\documentclass[journal,twoside,web]{ieeecolor}
\usepackage{generic}
\usepackage{cite}
\usepackage{amsmath,amssymb,amsfonts}
\usepackage{algorithmic}
\usepackage{graphicx}
\usepackage{textcomp}
\usepackage{amsmath}
\usepackage{amsfonts}
\usepackage{bm}
\usepackage{color}

\usepackage{bigstrut,multirow,rotating}
\usepackage{booktabs}
\usepackage{array}

\usepackage[caption=false, font=footnotesize]{subfig}

\def\BibTeX{{\rm B\kern-.05em{\sc i\kern-.025em b}\kern-.08em
    T\kern-.1667em\lower.7ex\hbox{E}\kern-.125emX}}
\begin{document}
\title{Joint Landmark and Structure Learning for Automatic Evaluation of Developmental Dysplasia of the Hip}
\author{Xindi Hu\textsuperscript{\dag}, Limin Wang\textsuperscript{\dag}, Xin Yang, Xu Zhou, Wufeng Xue, Yan Cao, Shengfeng Liu, Yuhao Huang, Shuangping Guo, Ning Shang*, Dong Ni*, and Ning Gu* 
\thanks{This work was supported in part by the National Key R\&D Program of China (2019YFC0118300), the Shenzhen Peacock Plan (KQTD2016053112051497, KQJSCX20180328095606003), and the Medical Scientific Research Foundation of Guangdong Province, China (B2018031).
}
\thanks{
X. Hu\textsuperscript{\dag} is with the School of Biomedical Engineering and Information, Nanjing Medical University, Nanjing, China.
}
\thanks{
L. Wang\textsuperscript{\dag}, S. Guo, and N. Shang* (e-mail: 499800208@qq.com) are with Ultrasound Department, Guangdong Women and Children Hospital, Guangzhou, China 
}
\thanks{
X. Yang, X. Zhou, W. Xue, S. Liu, Y. Cao, Y. Huang, and D. Ni* (e-mail: nidong@szu.edu.cn) are with the Medical UltraSound Image Computing (MUSIC) Lab, National-Regional Key Technology Engineering Laboratory for Medical Ultrasound, Guangdong Key Laboratory for Biomedical Measurements and Ultrasound Imaging, School of Biomedical Engineering, Health Science Center, Shenzhen University, Shenzhen, China.
}
\thanks{
N. Gu* (guning@seu.edu.cn) is with the School of Biomedical Engineering and Information, Nanjing Medical University, Nanjing, China; State Key Laboratory of Bioelectronics, Jiangsu Key Laboratory for Biomaterials and Devices, School of Biological Sciences and Medical Engineering, Southeast University, Nanjing, China.}
\thanks{
X. Hu and L. Wang contributed equally to this work.  }}

\maketitle 

\begin{abstract}
The ultrasound (US) screening of the infant hip is vital for the early diagnosis of developmental dysplasia of the hip (DDH). The US diagnosis of DDH refers to measuring alpha and beta angles that quantify hip joint development. These two angles are calculated from key anatomical landmarks and structures of the hip. However, this measurement process is not trivial for sonographers and usually requires a thorough understanding of complex anatomical structures. In this study, we propose a multi-task framework to learn the relationships among landmarks and structures jointly and automatically evaluate DDH. Our multi-task networks are equipped with three novel modules. Firstly, we adopt Mask R-CNN as the basic framework to detect and segment key anatomical structures and add one landmark detection branch to form a new multi-task framework. Secondly, we propose a novel shape similarity loss to refine the incomplete anatomical structure prediction robustly and accurately. Thirdly, we further incorporate the landmark-structure consistent prior to ensure the consistency of the bony rim estimated from the segmented structure and the detected landmark. In our experiments, 1,231 US images of the infant hip from 632 patients are collected, of which 247 images from 126 patients are tested. The average errors in alpha and beta angles are 2.221° and 2.899°. About 93\% and 85\% estimates of alpha and beta angles have errors less than 5 degrees, respectively. Experimental results demonstrate that the proposed method can accurately and robustly realize the automatic evaluation of DDH, showing great potential for clinical application.
\end{abstract}


\section{Introduction}
\label{sec:introduction}
\IEEEPARstart{D}{evelopment} dysplasia of the hip (DDH) is a common joint disease caused by the abnormal positional relation between the femoral head and the acetabulum, ranging from acetabular dysplasia to dislocation \cite{De2007Analysis}. According to the survey \cite{kang2017ultrasonography}, the incidence of DDH in newborns is approximately 1.5\textperthousand$\sim $2\%. The treatment of DDH in the neonatal period is simple and effective, with a success rate of 96\% \cite{kocer2016measuring}. If it can be diagnosed early, DDH-caused gait abnormalities, chronic pain, degenerative arthritis, and other long-term morbidities can be thus avoided \cite{shorter2013cochrane}.

\begin{figure*}[!t]
\centerline{
    \includegraphics[width=0.9\textwidth]{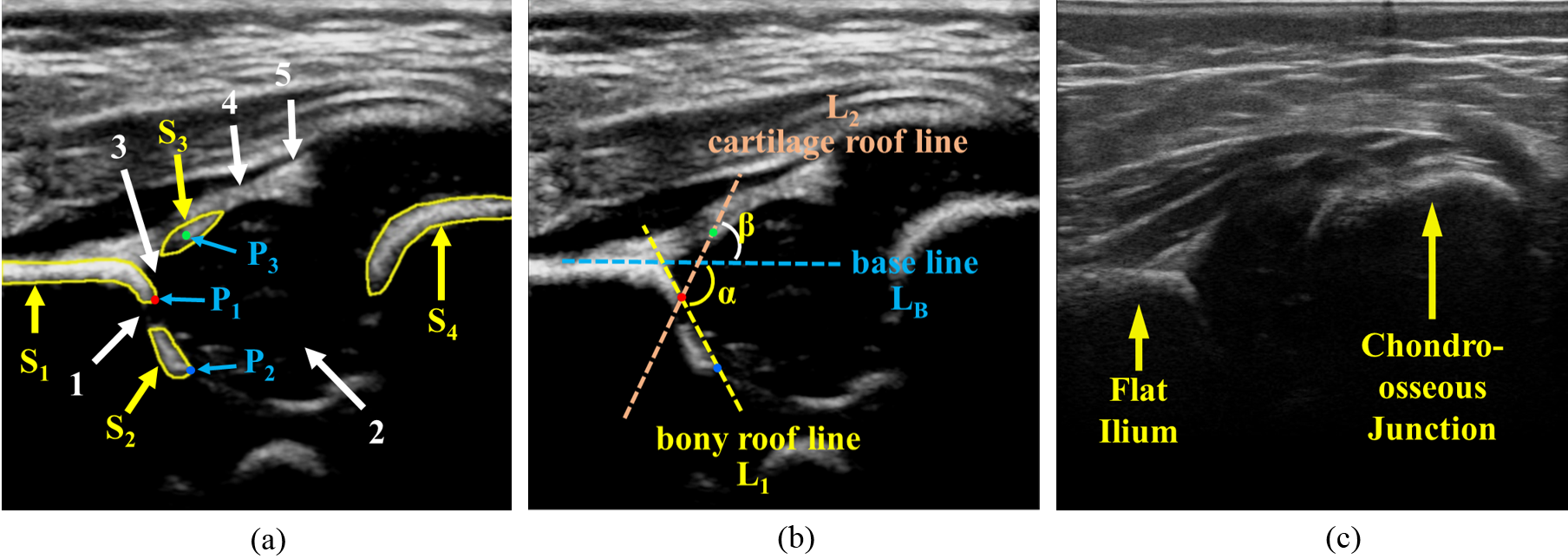}
    }
\caption{Examples of the hip joint. (a) An illustration for anatomical structures of the infant acetabulum, with 9 structures and 3 landmarks in total. Numbers 1-5 are common anatomical structures: 1, bony rim; 2, femoral Head; 3, hyaline cartilage preformed acetabular roof; 4, joint capsule; 5, synovial fold. Structure S1-S4 are key anatomical structures of the standard plane: S1, flat ilium; S2, lower limb; S3, labrum; S4, chondro-osseous junction. P1-P3 are three landmarks for the measurement: P1, bony rim point. It is the turning point of concavity to the convexity of the iliac bone. P2, lower limb point. It is on the floor of the acetabular fossa. P3, center of the labrum. (b) The measurement of two angles: alpha and beta. L1, base line. It is drawn caudally tangential to the flat ilium from the upper-most portion of the acetabular roof. L2, bony roof line. It is a tangent of the bony root which is drawn from the inferior rim of the ilium. L3, cartilage roof line. It is drawn from the bony rim through the center of the labrum. (c) A non-standard hip joint US image without lower limb and labrum.}
\label{DDHIntro}
\end{figure*}

In current clinical practice, ultrasound (US) is widely used for DDH screening and the Graf method \cite{graf2006hip} is one of the most established US examination techniques \cite{de2005hip}. Graf method consists of three steps: (1) identifying the standard plane (Fig. \ref{DDHIntro}(a)); (2) locating three landmarks in the identified plane; (3) measuring the main dysplasia metrics: $\alpha$ and $\beta$ angles. Fig. \ref{DDHIntro}(b) illustrates the definition of alpha and beta angles. The base line is drawn caudally tangential to the flat ilium from the upper-most portion of the acetabular roof. The bony roof line in yellow is the tangent of the bony root passing the lower limb point in blue. Coincidently, notice that it looks like the bony roof line passes the bony rim point in red, but not necessarily so by the definition. The cartilage roof line is drawn from the bony rim through the center of the labrum. $\alpha$ represents the angle between the base line and the bony roof line, quantifying the bony socket and the cartilage, and $\beta$ is the angle between the base line and the cartilage roof line, quantifying the cartilaginous acetabular roof.

Although the Graf method has been widely used clinically, its reliability is still controversial \cite{dias1993reliability}. Some adverse factors, such as imaging noise and artifacts, incomplete anatomical structures (Fig. \ref{DDHIntro}(c)), and high structural complexity of the hip joint (Fig. \ref{DDHIntro}(a)), impose a great challenge for doctors to make accurate measurements. It also results in high inter-observer variability in dysplasia metrics. Even measured by experienced clinicians, standard deviations of alpha and beta angles are as high as 3° and 6°, respectively \cite{quader2017automatic}. The variability is caused not only by the level of the physician but also by the quality of the US images \cite{wu2017fuiqa}. Therefore, a robust standardized and automated measurement method is highly demanded to reduce the variability.

In the past, there have been some studies assisting the clinical work of DDH through computer-aided diagnosis, which can be divided into traditional machine learning methods and deep learning methods. 

1) Traditional machine learning based methods mainly rely on manual feature extraction from US images\cite{4461934}. To segment the anatomy, different filters were adopted to improve the quality of hip joint US images. In \cite{kocer2016measuring}, the authors used the Geodesic Active Regions (GAR) model to segment the ilium. To measure the angles, Quader et al. \cite{quader2015automatic} proposed an automatic alpha and beta angle calculation system based on phase-symmetric features. They further used the same features to identify the bone and cartilage boundaries and then calculated key dysplasia metrics \cite{quader2017automatic}. \cite{hareendranathan2016toward} proposed a semi-automated method for the tracing of the bony surface. Alpha angle and rounding index were then measured automatically. Quader et al.\cite{quader2016improving} further extracted the ilium boundaries from US images with an isotropic bone feature extraction technique. Radon Transform of regions of interest (ROI) was then conducted around the inferior edge of the ilium to measure the alpha and the beta angles. However, due to the complexity of handcrafted feature design and engineering, traditional machine learning methods were limited in accuracy and robustness.

2) Deep learning based methods are being explored for intelligent DDH diagnosis because of the automated feature extraction capabilities of convolutional neural networks (CNNs)  \cite{8490669,7950734}. Deep learning methods for DDH analysis can be categorized as segmentation, angle measurement, and hip type classification. In terms of automatic segmentation of anatomical structures, El-Hariri et al. \cite{el2019comparative} proposed a deep-learned feature-based approach, which utilized the U-Net with multi-channel input to fulfill ilium segmentation. Their results showed that, compared to hand-crafted features, deep-learned features improved accuracy and speed, and reduced outliers and failure rates. Similarly,  Zhang et al. \cite{zhang2018end} introduced the ROI into the Full Convolution Network (FCN) to improve the segmentation of the acetabulum. To guide the segmentation, a multi-scale feature fusion network \cite{hareendranathan2017toward} was presented to output a probability map of the bone. Additionally, Sezer et al. \cite{7950680} incorporated global and local features into a CNN framework to obtain a probability map of the bone and then used this map in probabilistic graph search to guide the segmentation. In terms of automated angles measurement, the adversarial idea \cite{golan2016fully} was used in the CNN to segment the flat ilium and lower limb, resulting in 77\% alpha measurement errors within 5°. In addition, there are methods that aim to classify hip joints. In \cite{sezer2020deep}, authors used a region-based active contour model to segment the iliac wing in the defined ROI of US images. The segmented image patches were input into a CNN to directly diagnose the infant hip as normal, mild, or severe dysplasia. Hareendranathan \cite{hareendranathan2021impact} proposed a 10-point scoring system to evaluate the quality of DDH US scan, which is based on multiple anatomical structures.

Although the above studies have made important contributions to the diagnosis of DDH, there are still two main shortcomings: 1) the neglect of the standard plane, and 2) the insufficient utilization of information on anatomical structures.
The existing methods only handle part of the anatomical structures, without considering the anatomical information of all 4 structures and 3 landmarks. Therefore, the lack of an overall analysis of the anatomies in the hip joint images may lead to poor performance.

\begin{figure*}[!t]
\centerline{
    \subfloat[]{\includegraphics[width=0.3\textwidth]{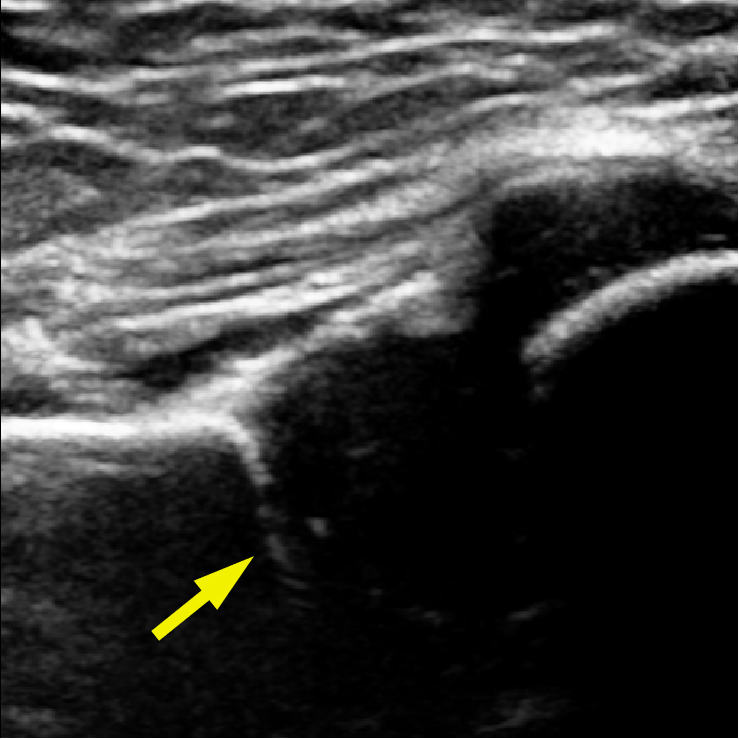}}
    \hspace{1mm}
    \subfloat[]{\includegraphics[width=0.3\textwidth]{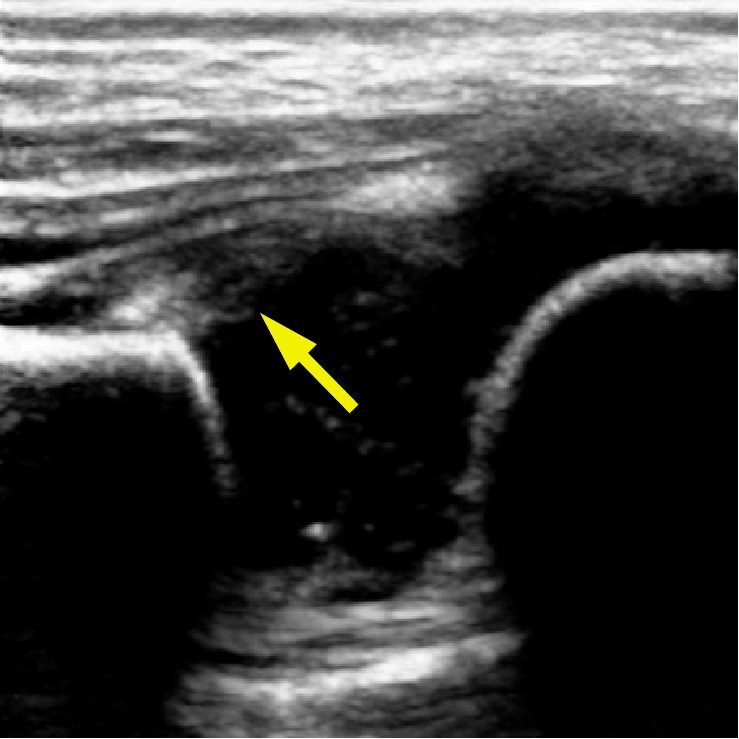}}
    \hspace{1mm}
    \subfloat[]{\includegraphics[width=0.3\textwidth]{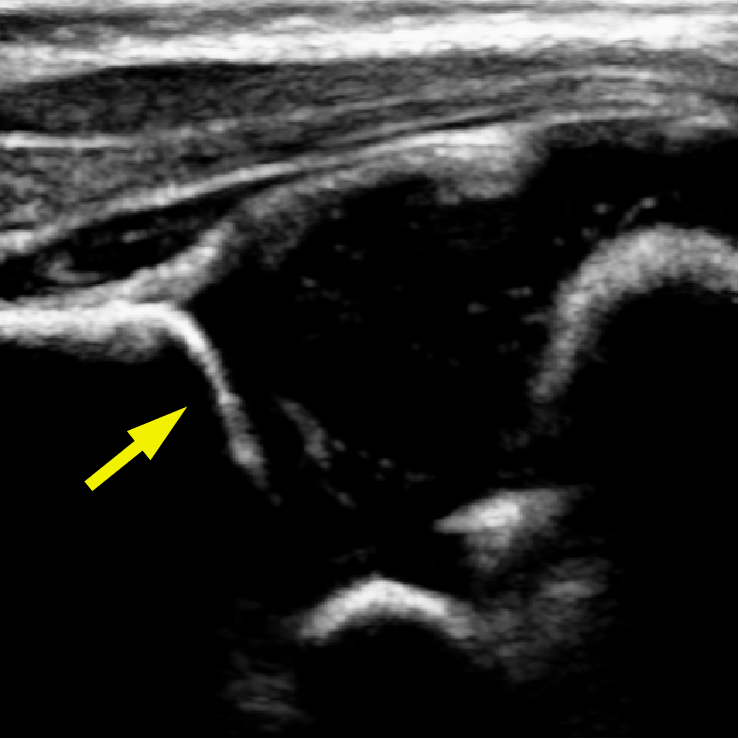}}
    }
\caption{Three hard examples for segmentation. (a) The indistinct lower limb. (b) The invisible outline of the labrum. (c) Due to the connection of the flat ilium and the lower limb, there is no hypoechoic at the bony rim. The yellow arrow pinpoints the problematic sites.}
\label{Challenges}
\end{figure*}

The judgment of the standard plane is an indispensable part of DDH diagnosis, but it is often ignored by existing intelligent diagnosis methods. As the baby moves and the scanning angle changes, some structures may not be fully displayed on the US image. For example, although the chondro-osseous (CO) junction does not affect the measurement angles, the absence of CO junction indicates that the current image is not a standard plane and hence worthless for diagnosis. The identification of anatomical structures often needs extensive experience in DDH diagnosis and is a non-trivial prerequisite step for landmark detection. If the process of identifying anatomical structures can be completed in an automated and standardized manner, it can not only greatly reduce the burden on sonographers, but also guide inexperienced physicians. Therefore, in our method, we leverage segmentation information of four key anatomical structures for identifying standard planes, and in turn to ensure that the subsequent measurement is efficient.

The measurement of dysplasia metrics depends on the shape information of the hip joint anatomy, which can be effectively obtained by segmentation. However, there are two major challenges to segment the hip joint from the US image: 1) the segmentation of incomplete anatomical structures due to imperfect standard planes poses a great difficulty. Blurry structural edges (such as low echo of the lower limb in Fig. \ref{Challenges}(a) and the invisible inferior curve of the labrum (Fig. \ref{Challenges}(b)) make the accurate segmentation challenging. The invisible structures often lead to fragmental segmentations and subsequent inaccurate angle estimates. 2) the segmentation of connected anatomical structures, including but not limited to the flat ilium and lower limb is also difficult. As shown in Fig. \ref{Challenges}(c), the boundary between the two structures is difficult to be distinguished, resulting in inaccurate localization of bony rim and estimation of beta angle.

Multi-task learning can well improve the generalization ability of learning by sharing feature information between different tasks~\cite{caruana1997multitask,2015Deep}. In deep learning, multi-task learning is usually completed by sharing hard or soft parameters of the hidden layer. For hard sharing of parameters, the usual method is to share convolutional layers while learning fully connected layers for specific tasks. For example, the Deep Relationship Networks proposed by \cite{2015Learning} and the Fully-Adaptive Feature Sharing Networks proposed by \cite{Lu_2017_CVPR}. But this method has proved to be error-prone to new and complex tasks. Soft sharing of parameters represents connecting several independent networks through soft parameters to increase learning ability. For example, Cross-Stitch Networks proposed by \cite{2016Cross} and the weighting losses with uncertainty methods proposed by \cite{2018Multi}. It is proved to be effective by adjusting the relative weight of each task in the loss function\cite{zhang2014facial,9369104,8363713}.

In this work, different from the existing multi-task learning frameworks, we take into account the spatial relationships among anatomical structures and landmarks and propose a multi-task learning network that is more suitable for DDH measurement and diagnosis. Our multi-task networks are equipped with three novel modules to implement automatic and accurate angle measurements while marking key anatomical structures. Quantitative results on 247 images from 126 patients show that the average errors in alpha and beta angles are 2.221° and 2.899°, with 93\% and 85\% estimation errors less than 5 degrees, respectively. It demonstrates that our proposed multi-task network has great potential for clinical application. The contributions of our method can be summarized in threefold.

1) Our multi-task learning network can accurately accomplish anatomical structures detection and segmentation, and landmark detection. Mask R-CNN \cite{he2017mask} is adopted as the basic framework for the detection and segmentation of four anatomical structures. A landmark detection branch is inserted to localize three key landmarks. 

2) We leverage priors of anatomical structural shape into the network by following the spirit of the curve similarity \cite{yan2017skeletal} to deal with the challenges of fragmental segmentation. We explore it as a similarity loss function to regularize the borders of segmentation so that an intact mask can be recovered from an incomplete anatomical structure.

3) We model the landmark-structure consistent prior as a novel loss function, called bony rim loss, for segmenting the connected structures. It can enforce the bony rim estimated from the mask of the flat ilium to be consistent with the detected landmark so that the touched anatomical structure can be correctly segmented.

\begin{figure*}[!t]
\centerline{
    \includegraphics[width=\textwidth]{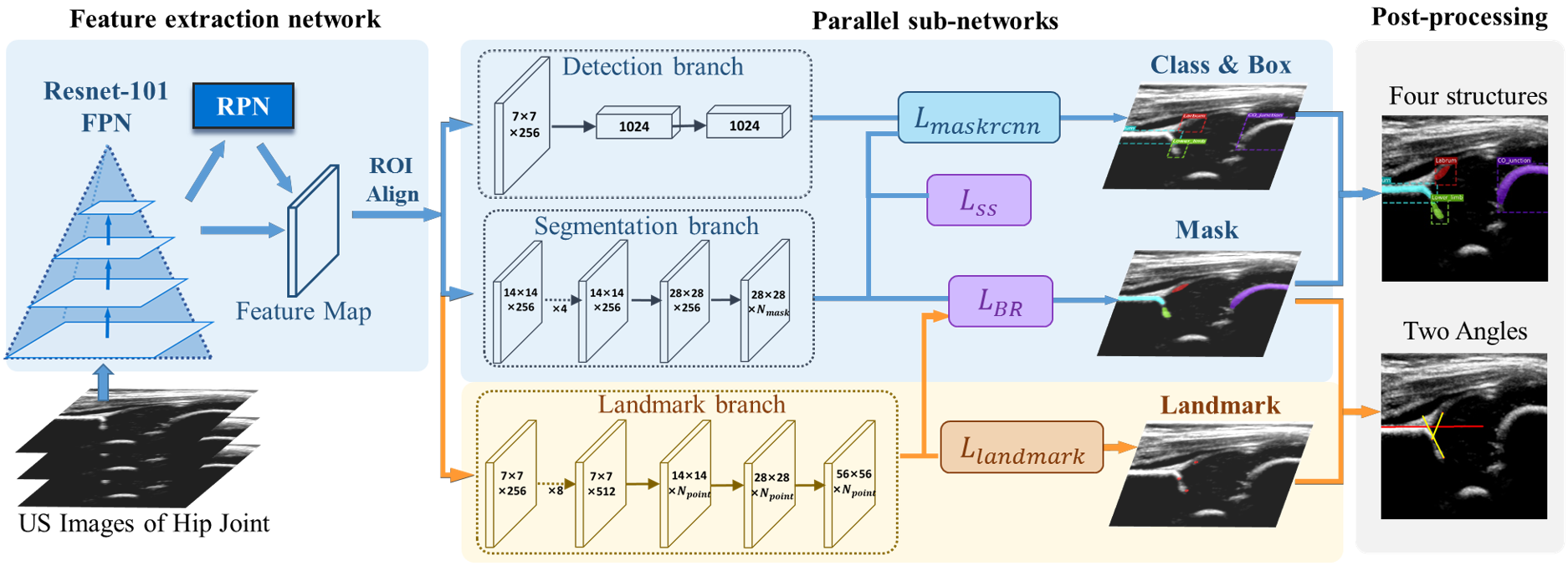}
    }
\caption{The pipeline of the proposed framework. The blue region is Mask R-CNN as the model backbone involving the loss~($L_{maskrcnn}$). The yellow region is a landmark detection branch involving the loss~($L_{landmark}$) and the purple module represents additional structure priors involving shape similarity loss~($L _{SS}$) and bony rim loss~($L _{BR}$).}
\label{Network}
\end{figure*}

\section{Methods}
\label{sec:methods}
Fig. \ref{Network} shows the pipeline of our proposed method. US images of the hip joint are input into the network and forwarded by a feature extraction network. The extracted feature maps are then fed into three parallel sub-networks for structure detection, segmentation, and landmark prediction tasks, respectively. Four structures in the form of the mask are obtained from the detection and segmentation tasks. The preliminary positions of three landmarks are predicted from the landmark detection branch. Next, we obtain the final positions of three landmarks by combining the location information of two landmarks, which are inferred from the segmentation mask and predicted from the landmark detection branch, respectively. Finally, the DDH metrics including alpha and beta angles are calculated based on estimated masks and detected landmarks.

\subsection{Multi-task Learning Network}
To take advantage of the spatial information between anatomical structures and landmarks, we propose a multi-task learning network that combines subtasks, including detection, segmentation of multiple anatomical structures, and prediction of multiple landmarks.

As the main basis for DDH measurement, the segmentation of anatomical structures requires a network framework with excellent performance. Recently, popular segmentation networks, such as U-net \cite{ronneberger2015u} and FCN \cite{long2015fully}, have been used to segment the hip image. It is also well known that cropping out the area containing the target structure can eliminate the interference of other structures, and the segmentation will be greatly improved \cite{sezer2020deep}. It is similar to the instance segmentation in \cite{bolya2019yolact,Xie2019PolarMaskSS,2017Hierarchical,2019CT,2021Diverse}. In this study, we adopt Mask R-CNN, which can simultaneously perform detection and segmentation tasks, as the basic network.

Although the positioning of three landmarks can be roughly estimated by an end-to-end landmark detection network\cite{2019Context,2018Anatomical,2017Alzheimer,2017Landmark,2020Automatic}, the detection performance may be degraded because of the large appearance variabilities of landmarks and relatively low quality of US images. The purpose of the multi-task is to allow the network to additionally learn the potential associated information of different structures in the image so that the positioning of landmarks and the segmentation of anatomical structures can be improved.

Our network is divided into two stages. The first step is to extract multi-scale features of the input images via the backbone, which is ResNet-101-FPN in our work. The output feature maps are then fed into the Region Proposal Network (RPN)\cite{2017Faster} to determine the bounding boxes of the candidate regions. In the second step, the extracted feature maps go through a layer for extracting a small feature map from each Region of Interest, called RoIAlign, and the feature map is then assigned to multiple subtasks for target detection, segmentation, and landmark detection. The specific implementation of the first two tasks and the loss function $L_{maskrcnn}$ composed of classification $L_{cls}$, boundary box regression $L_{box}$ and segmentation $L_{mask}$, are based on the work in \cite{he2017mask}. The loss function for landmark detection is softmax cross-entropy loss defined as:
\begin{equation}
L _{landmark}=-\sum_{i=1}^{N}y_{i}log(y_{i}^{,}),
\label{LandmarkLoss}
\end{equation}
where $y_i$ is the $i$-th landmark label (0 or 1) and ${y}'_i$ is the prediction probability of the $i$-th label normalized by softmax.

\subsection{Prior on Shape Similarity}

\begin{figure*}[!t]
\centerline{
    \includegraphics[width=0.95\textwidth]{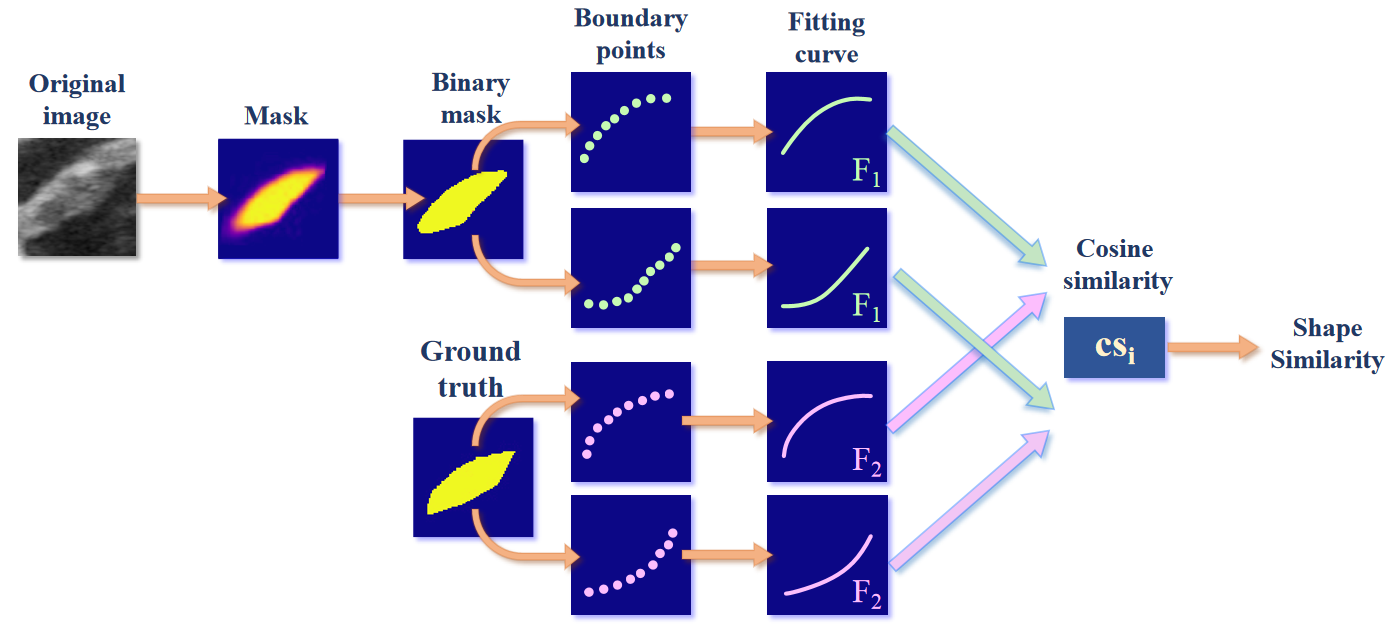}
    }
\caption{The pipeline of our proposed SS loss quantifies the shape similarity.}
\label{SSFig}
\end{figure*}

As for the challenge that incomplete anatomical structures may lead to fragmental segmentations in the US image of the hip joint, as shown in Fig. \ref{Challenges}(a) and Fig. \ref{Challenges}(b), we propose to introduce the prior knowledge of anatomical structure to alleviate it. Since the shape of the same anatomical structure usually has a certain similarity, the shape information can be input into the network as the priors so that the network can still output the segmentation results that match the original shape in the case of partially missing anatomical structures.

Inspired by the curve similarity presented in \cite{yan2017skeletal}, we propose a novel Shape Similarity (SS) loss to regularize the network and use the reference shape of the anatomical structure to improve the segmentation, see Fig. \ref{SSFig} for the pipeline. The binary prediction map is the output by the mask branch at first. Since the segmentation boundary is usually a closed curve, we divide it into two curves, such as the upper and the lower curves. The boundary of structure $j$ can then be approximately fitted with two cubic polynomials
\begin{equation}
C_{j}\left ( x \right )=a_{j}x^{3}+b_{j}x^{2}+c_{j}x+d_{j},
\label{Eq:CurveSimilarity}
\end{equation}
where $a$ to $d$ are the coefficients of the equation. We use the least-squares method to find the coefficients of this equation and the solution is in the closed-form of
\begin{equation}
\theta =\left ( X^{T}X+\xi I \right )^{-1}X^{T}X,
\label{Theta}
\end{equation}
where $X$ and $Y$ are the abscissas and ordinate matrix of points on the curve, $\xi$ is the regularization weight, and $I$ is the identity matrix. The coefficients of (\ref{Eq:CurveSimilarity}) other than the constant term are extracted as the coefficient vector $F=\left [a,b,c \right ]$ .

For a boundary of an anatomical structure, we obtain the coefficient vector $F_1$ from the mask and the coefficient vector $F_2$ from the ground truth by (\ref{Eq:CurveSimilarity}) and then calculate the cosine similarity of these two vectors by
\begin{equation}
cs_{i}=\begin{vmatrix}
\frac{<F_{1},F_{2}> }{\left |F_{1} \right |,\left |F_{2}  \right |}
\end{vmatrix},
\label{CosineSimilarity}
\end{equation}
where $\left \langle \cdot  \right \rangle$ means the dot product operation and $\left | \cdot   \right |$ means the length of the vector. If the two vectors tend to be similar, the value of $cs_i$ will be close to 1, otherwise, it is close to 0. Therefore, we define our SS loss on four structures of the hip joint as
\begin{equation}
L_{SS}=\sum_{i=1}^{N}\sum_{j=1}^{4}\left ( 1-cs_{ij} \right ).
\label{SSLossFunction}
\end{equation}

SS loss encodes the shape prior by emphasizing the edge-wise consistency between the predicted mask and ground truth. During network training, $L_{SS}$ penalizes the large segmentation discrepancy which is often caused by the blurry edges or incomplete structures in hip joint US images. Our network is hence able to recover boundary predictions and alleviates the failure cases.

\subsection{Prior on Landmark-structure Consistent}
In the hip US image, three landmarks are essential for measuring angles, among which the bony rim is the most difficult to locate. The bony rim is located at the critical line where the low rim of the flat ilium transitions from hyperechoic to hypoechoic. However, in clinical practice, sometimes the flat ilium and lower limb are connected, resulting in the absence of a hypoechoic area, as shown in Fig. \ref{Challenges}(c). In this case, it is important for the network to learn the positional relationship between the bony rim and the flat ilium, so that the bony rim could fall on the edge of the flat ilium segmentation as much as possible. We introduce the landmark-structure consistent prior to the network which enables the mask branch and the landmark branch to supervise each other and achieve better consistency. 

The bony rim can be obtained by two methods from our network: one is inferred from the segmentation mask by the definition that the bony rim is located on the tail of the lateral edge of the flat ilium; the other is predicted by the landmark branch. The results of the two methods should be the same point. In other words, the distance between these two points is 0. We modify this distance as a loss function, named BR loss, to constrain both branches by extracting bony rim from the mask branches and the landmark branch respectively.

In the process of calculating BR loss, the coordinates of two bony rim landmarks obtained from the structure segmentation and the landmark detection are denoted as $m _{i}$ and $k _{i}$, respectively. Given the number of samples, we calculate the Euclidean distance between two points. The BR loss is defined as the $L_{2}$-norm given by
\begin{equation}
L_{BR}=\sum_{i=1}^{N}\sqrt{\sum_{j=1}^{2}\left (m_{ij}-k_{ij} \right )^{2}}.
\label{BRLoss}
\end{equation}

\subsection{DDH Measurement}


Our method automatically measures the diagnostic indicators of DDH in the prediction phase. The segmentation branches output four masks: flat ilium $S_1$, lower limb $S_2$, labrum $S_3$, and CO junction $S_4$; the landmark branch outputs primary positions of three landmarks: the bony rim $P_{k1}$, the lower limb point $P_{k2}$ and the labrum point $P_{k3}$. Taking into account the close correlation between the landmark and anatomical structure (see Fig. \ref{DDHIntro}(a)), landmark positions predicted by the two strategies are averaged for better accuracy and robustness. The final positions of three landmarks are calculated by the formula
\begin{equation}
P_{i}=\left (P_{si}+P_{ki}  \right )/2,\: \left ( 1\leq i\leq 3 \right )
\label{PointPosition}
\end{equation}
where $P_{si}$ is the landmark inferred from segmented structures. 

With the three landmarks in hand, we can automatically measure the angles: alpha and beta (see Fig. \ref{DDHIntro}(b)). The base line $L_B$ is the tangent of the flat ilium; the bony roof line $L_1$ is the tangent line to the outer contour of flat ilium at the lower limb point; the cartilage roof line $L_2$ is the line that connects the bony rim to the center of the labrum. The alpha angle is calculated between $L_B$ and $L_1$, and the beta is calculated between $L_B$ and $L_2$.

The final loss function of our method is defined as
\begin{equation}
L=\lambda _{1}L _{maskrcnn}+\lambda _{2}L _{landmark}+\lambda _{3}L _{BR}+\lambda _{4}L _{SS},
\label{TotalLoss}
\end{equation}
where $\lambda$ is regularization weight. Fig. \ref{Network} shows the overall network architecture as well as the losses for each subtask.

\section{Experiment}
\label{sec:Experiment}
\subsection{Data Preparation}
\subsubsection{Data Collection}
In our experiment, we collect the hip joint US images of 632 infants at the age ranging from 0 to 6 months. Each patient contains one to two images of the left and right legs, resulting in a total of 1,231 images. 
These images consist of 1070 type 1 hip joint and 161 type 2 hip joint cases without dislocation. Under the approval of the Institutional Review Boards, all the images are acquired from two machines of Guangdong Women and Children Hospital, and the models of which are Hitachi HI-Vision Preirus (Machine 1) and Philips iU22 (Machine 2), respectively. 
All US images and annotations in our dataset have undergone strict quality control to guarantee the reliability of the ground truth for the model training. Firstly, each image is reviewed by a 15-year experienced expert to ensure that it is a clinically usable standard image.
All images are manually labeled by an experienced doctor, and all the labels are finally reviewed and inspected by a group of senior doctors. The labels include four segmented structures (flat ilium, lower limb, labrum, and CO junction) and three landmark points (bony rim, lower limb point, and the midpoint of the labrum). 
Note that each image has two measurements from 2 different doctors: one from the doctor engaged in labeling and the other from the measurement during clinical diagnosis.
In our work, the one from manual labels (including the contour of 4 key anatomical structures and the position of 3 landmarks) is the ground truth, and the one from the manual plot (alpha and beta angles) during clinical diagnosis is only used for calculating the disagreement in angle measurement between doctors.

\subsubsection{Data Distribution}
 We randomly split the data into 506 and 126 cases for training and testing at the patient level to ensure that multiple images of one patient belong to the same set. The detailed data distribution is shown in Table \ref{DataTable}. In addition, standard five-fold cross-validation is also utilized, of which four-fold is used for the training set and the rest for the testing set. Besides, there are three pre-processing steps before an image goes into our framework. First, the key image area is cropped from the original US image to prevent invalid information from affecting the network training. Second, all the images are resized to 512×512 and are augmented using rotation, translation, contrast enhancement, and brightness transformation. The last step is standardizing all the image intensities to 0-1. The above pre-processing is completed in a fully automatic way without manual factors.

\begin{table}
\caption{Detailed Data Distribution. Digits Inside and Outside of Parentheses Represent Image and Case Numbers, Respectively.}
\centering
\def\temptablewidth{1\columnwidth}
\begin{tabular*}{\temptablewidth}{@{\extracolsep{\fill}}cccc}
   \toprule[1pt]
    Machine Types & Sum & Training set & Testing set \\
    \midrule
    Machine 1 & 268 (584) & 210 (475) & 58 (110) \\
    Machine 2 & 364 (647) & 296 (510) & 68 (137) \\
    Sum & 632 (1231) & 506 (985) & 126 (247) \\
    \bottomrule[1pt]
\end{tabular*}
\label{DataTable}
\end{table}

\subsection{Experimental Evaluation}
\subsubsection{Experimental Setup}
In this study, we evaluate the method from two perspectives. First, we evaluate the measurement performance by dysplasia metrics calculated by the difference between the predicted and the manual angles. We also calculate the error rates for classifying type I and type II hip joint. Second, we evaluate the task performance of structure segmentation and landmark localization and also evaluate the efficacy of the proposed SS loss and BR loss. We set up a total of six experiments: (a) Assessment of dysplasia metrics and the classification of the hip joint. (b) Segmentation of four anatomical structures. (c) Prediction of three landmarks. (d) Efficacy of BR loss. (e) Influence of regularization weights. (f) System performance on image quality, different machines, and hip types.

Additional significant test experiments are conducted to demonstrate the efficacy of our method. We used the two-sample t-test and independently repeated 10 experiments. The significance level is 0.05 and the degree of freedom (DF) is 18, and the p-value less than the significance level indicates a significant difference between the two population means.

\subsubsection{Evaluation Metrics}
We utilize the Dice Similarity Coefficient (DSC) to evaluate the overall segmentation performance to measure the similarity between prediction and ground truth, which is expressed as
\begin{equation}
DSC\left ( X,Y \right )=\frac{2\left | X\bigcap Y \right |}{\left | X\right |+\left | Y\right |},
\label{DSCEq}
\end{equation}
where $Y$ denotes the points set of the prediction mask and X denotes that of the ground truth. Since segmented contours have a great influence on the landmark locations, we also evaluate the quality of the segmentation edge via Hausdorff Distance (HD) that is defined in \cite{232073} as
\begin{equation}
H\left ( X,Y \right )=\left ( \max_{x\in X}\min_{y\in Y}\left \| x-y \right \|,\max_{y\in Y}\min_{x\in X} \left \|y-x \right \|\right ),
\end{equation}
where $\left \| \cdot \right \|$ is the $L_{2}$ distance of points of $x$ and $y$.

\subsubsection{Implementation Details}
In our work, we use ResNet-FPN-101 as the backbone. The loss weights $\lambda_1$ and $\lambda_4$ are default to 1, and $\lambda_2$ and $\lambda_3$ are set to 0.5 in (\ref{TotalLoss}). The size of the ROI is set to $7\times7$ in the landmark branch. Other parameters are consistent with \cite{he2017mask}. The model is implemented in the TensorFlow framework and trained on one NVIDIA GTX 2080Ti GPU in Linux with the Inter Xeon Silver 4114 CPU @2.20GHz. The weights of Mask R-CNN pre-trained on the COCO dataset are firstly imported. The network head for bounding-box recognition, mask prediction, and landmark detection is trained for 20 epochs with a learning rate of $1\times10^{-3}$, and all layers are finetuned for 250 epochs with a learning rate of $1\times10^{-4}$. The whole training process costs about five hours.

\section{Results}
In all experiments, "Unet", "PSPnet" and "DeeplabV3+" represent three advanced segmentation methods proposed in  \cite{2015U,2016Pyramid,2018Encoder}, respectively. "Mask R-CNN" stands for the original Mask R-CNN proposed in \cite{he2017mask}, which includes instance detection and segmentation, and the positioning of landmarks is determined only by the segmentation results. "+ SS loss" means that SS loss is added to train the original Mask R-CNN. "+ Landmark" means that the landmark detection branch is incorporated into the Mask R-CNN with SS loss added in training, and primary landmark positions are directly predicted by the landmark branch. "Our method" is the Mask R-CNN with the landmark branch and the training loss in (\ref{TotalLoss}), and the ultimate landmark position is the average of segmentation and landmark detection, as shown in (\ref{PointPosition}). Asterisk (*) denotes significant differences when other methods are compared with our method, and the performance of our method has a significant improvement over other methods. Bold in the table indicates it has the best performance compared with other methods.

\begin{table}
  \def\temptablewidth{1\columnwidth}
  \caption{Mean Absolute Difference and Standard Deviation of $\alpha$ and $\beta$ Angle (° ). The Asterisk (*) Denotes Significant Differences.}
    \begin{tabular*}{\temptablewidth}{@{\extracolsep{\fill}}cll}
    \toprule[1pt]
     & \multicolumn{1}{c}{Angle $\alpha$ (std) (p-value)} & \multicolumn{1}{c}{Angle $\beta$ (std) (p-value)} \\
    \midrule
     Manual plot & 3.984* (3.563) (p\textless0.01) & 5.831* (4.662) (p\textless0.01)\\
     Unet &  3.815* (9.401) (p\textless0.01)  &  7.162* (8.571) (p\textless0.01)  \\
     PSPnet & 2.503* (2.694) (p\textless0.01)  & 5.364* (10.764) (p\textless0.01)  \\
     DeeplabV3+ & 2.348* (2.587) (p=0.027)  & 3.276* (3.717) (p\textless0.01)  \\
     Mask R-CNN & 2.762* (3.573) (p\textless0.01)  & 3.220* (2.493) (p\textless0.01) \\
     + SS Loss & 2.589* (2.267) (p=0.014)  & 3.116* (2.399) (p=0.019)  \\
     + Landmark & 2.499* (3.179) (p\textless0.01)  & 3.055* (2.409) (p=0.032)  \\
     Our Method  & \textbf{2.221} (2.007)  & \textbf{2.899} (2.283) \\
    \bottomrule[1pt]
    \end{tabular*}%
\label{AngleResult}%
\end{table}%

\begin{figure}
\centerline{
    \begin{minipage}[b]{\columnwidth} 
    \includegraphics[width=0.9\columnwidth]{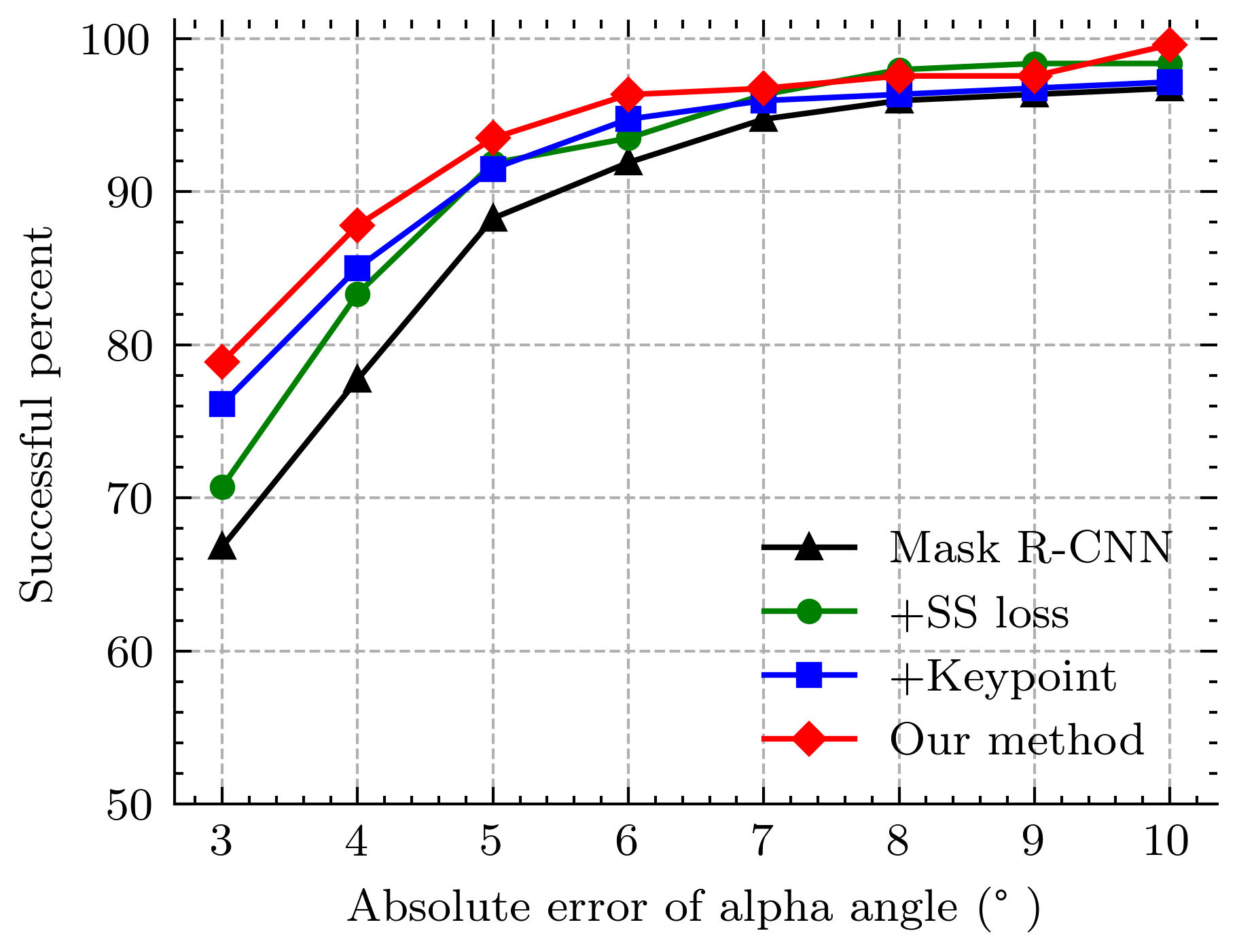}\\
    \includegraphics[width=0.9\columnwidth]{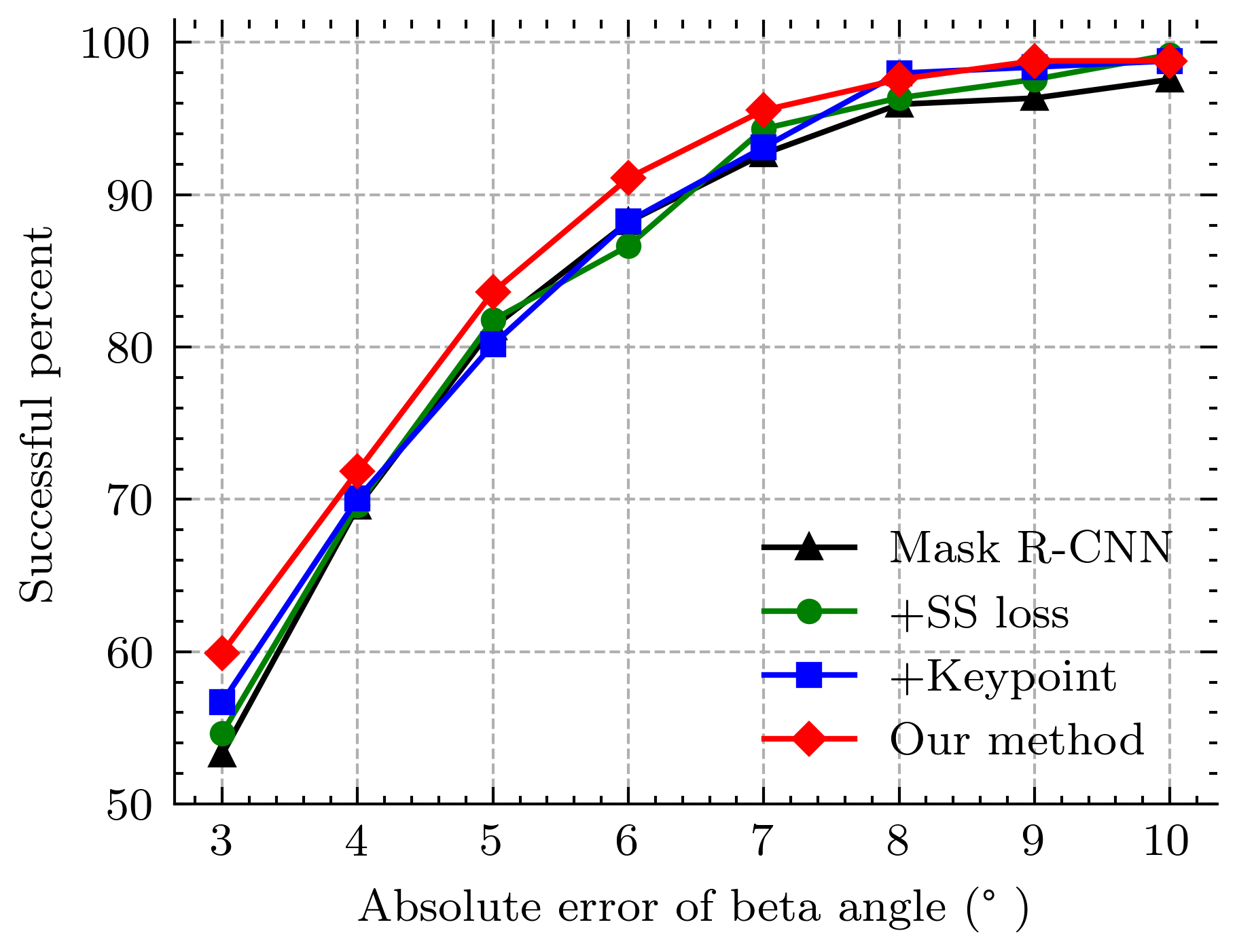}
    \end{minipage}}
\caption{The cumulative distribution of absolute errors concerning alpha and beta angles.}
\label{ErrorHistogram}
\end{figure}

\begin{figure*}
\centerline{
    \centering
    \includegraphics[width=0.85\textwidth,height=0.73\textwidth]{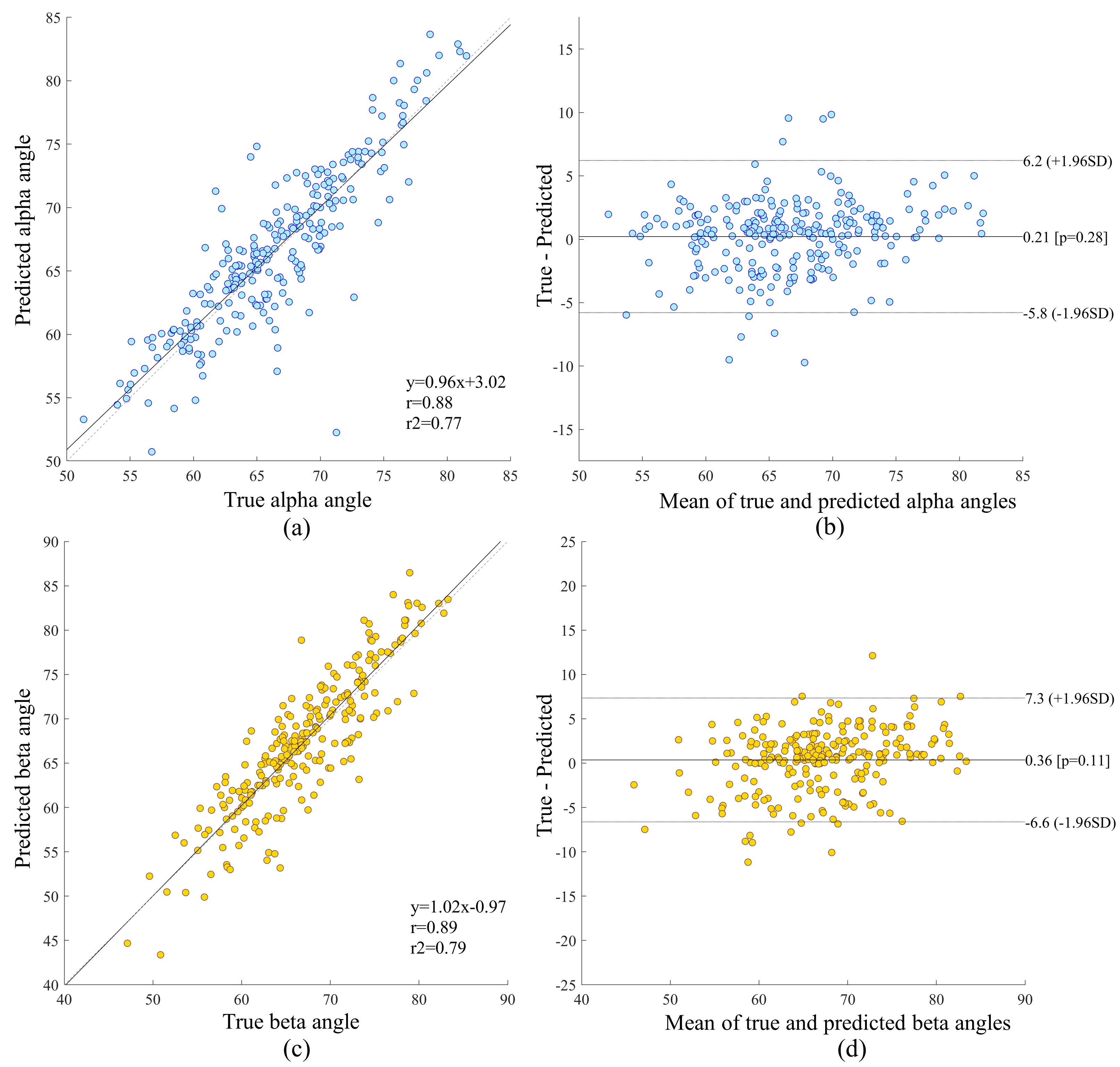}
    }
\caption{Scatter plots of the predicted and true values of the alpha and beta angles. (a)(b) represent the comparison between the real value and the predicted value of the alpha angle. (c)(d) represent the comparison between the real value and the predicted value of the beta angle.}
\label{Scatterplots}
\end{figure*}

\subsection{Assessment of Dysplasia Metrics}
Table \ref{AngleResult} shows the mean and standard deviation (std) of angle errors between the predicted and the manual annotations. 
Thus, our method significantly performs better than manual plots by doctors and various other methods (Unet, PSPnet and DeeplabV3+). It reveals the efficacy of our multi-task learning framework for sharing feature information among different tasks.
Besides, the average error is consistently reduced as more proposed components being incorporated into the Mask R-CNN. This verifies that our proposed techniques can well leverage the shape priors and landmark knowledge, which can help improve the system performance. After the addition of both losses and the operation of averaging landmarks obtained from the landmark and the mask branches, our method outperforms the Mask R-CNN significantly in terms of mean accuracy and stability. Especially, it achieves an improvement of 0.521 and 1.556 degrees in mean and standard deviation for the alpha angle.

The cumulative distributions of predicted angle errors are shown in Fig. \ref{ErrorHistogram}. It can be seen that our method is the best in terms of mean accuracy, with 93.456\% and 84.615\% successful angle estimates for alpha and beta angles. Here we suppose that angle estimate with an error of fewer than 5 degrees are considered successful since it is close to a professional doctor’s performance. We also report that the proposed method achieves 0.407\% and 1.215\% poor estimates for alpha and beta angles (we consider that an angle estimate with an error of more than 10 degrees is poor), whereas the Mask R-CNN has poor rates of 3.239\% and 2.444\% for alpha and beta angles. In short, it can be concluded that the proposed method significantly improves accuracy and robustness.

Scatter plots of the predicted and true values of the alpha and beta angles are shown in Fig.~\ref{Scatterplots}. 
The Pearson correlation coefficient of the alpha and beta angles are 0.88 and 0.89, indicating that there is a strong correlation between the predicted and true values of the two angles. 
 In addition, the two-sample t-test is used to compare the difference between the two values. The p-values of the alpha and beta angle are 0.28 and 0.11, respectively, with both of them larger than 0.05. Therefore, there is no significant difference between the predicted and the true value of the two angles.

\begin{table}
  \centering
  \def\temptablewidth{1\columnwidth}
  \caption{The Miscalassication Rate of the Hip Joint Category (\%).The Asterisk (*) Denotes Significant Differences.}
    \begin{tabular*}{\temptablewidth}{@{\extracolsep{\fill}}clll}
    \toprule[1pt]
    & Overall errors& FN & FP  \\
    \midrule
     Mask R-CNN & 10.121* & 8.097* & 2.024* \\
     Our Method  & \textbf{5.668}  & \textbf{4.453}  & \textbf{1.215} \\
    \bottomrule[1pt]
    \end{tabular*}%
    \label{DDHClass}%
\end{table}%

\begin{figure*}[!t]
\centerline{
    \includegraphics[width=0.75\textwidth,height=0.55\textwidth]{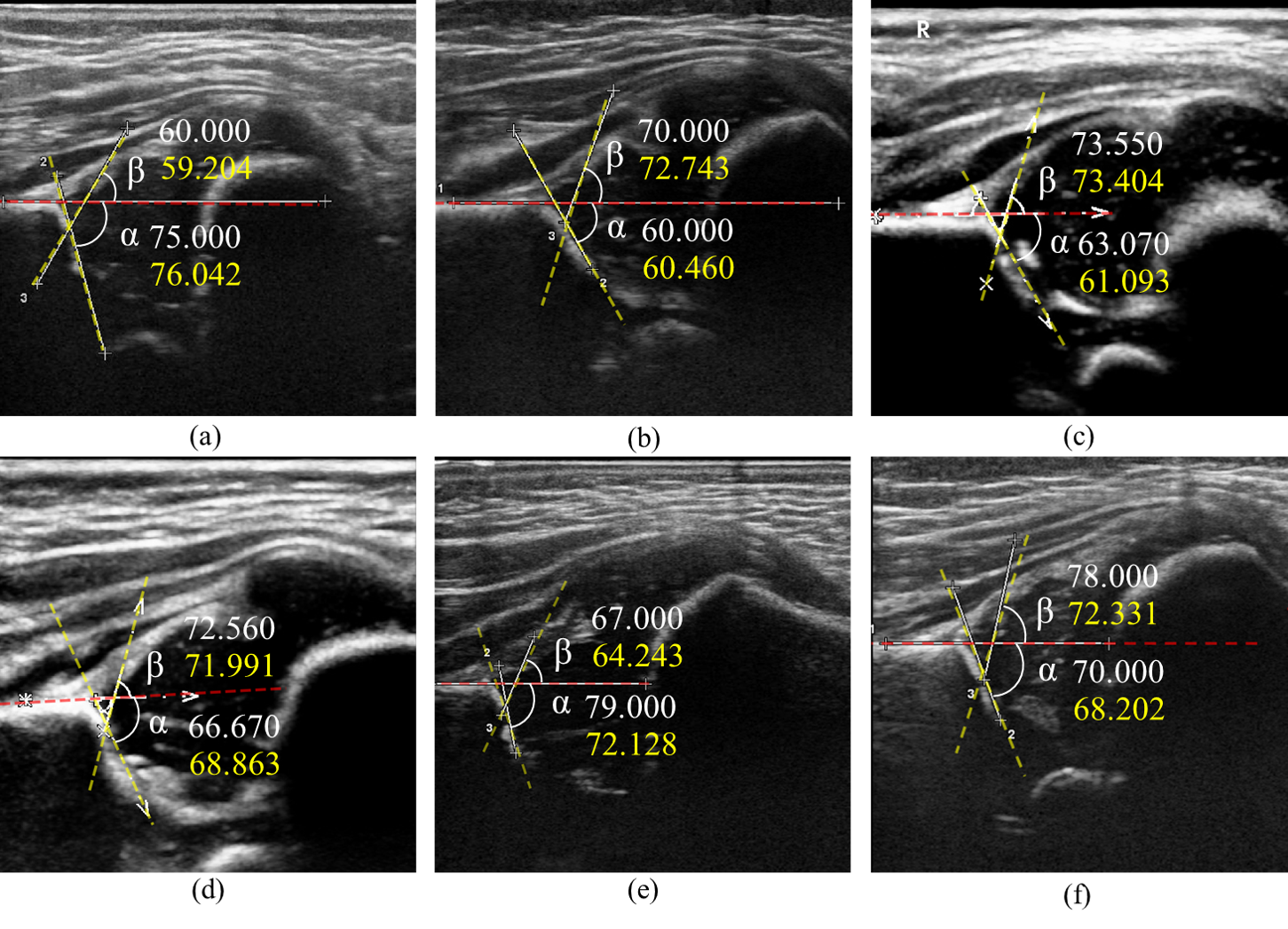}
    }
\caption{Qualitative comparisons on the angle measurement between our method and manual annotations. The white lines are the manual plot line in the clinic. The red and yellow lines are predicted by our method. The white value is the manual angle plot, and the yellow is the prediction. The unit is degree. The absolute differences between our method and the manual plot are (a) $\Delta\alpha = 1.042^{\circ}$, $\Delta\beta = 0.796^{\circ}$. (b) $\Delta\alpha = 0.460^{\circ}$, $\Delta\beta = 2.743^{\circ}$. (c) $\Delta\alpha = 2.193^{\circ}$, $\Delta\beta = 0.569^{\circ}$. (d) $\Delta\alpha = 1.977^{\circ}$, $\Delta\beta = 0.146^{\circ}$.(e) $\Delta\alpha = 6.872^{\circ}$, $\Delta\beta = 2.757^{\circ}$. (f) $\Delta\alpha = 1.798^{\circ}$, $\Delta\beta = 5.669^{\circ}$.}
\label{AngleComparison}
\end{figure*}

In clinical practice, doctors focus more on the alpha angle because it determines the type of the hip joint, while the beta angle only determines the subclass. To prove the clinical effectiveness of our method, we evaluate the classification performance of automatic measurement on Graf’s categories. The hip joint images are classified into two categories based on the predicted alpha angle, considering that there are only two types of hip joints (type I and type II) in our dataset. The classification errors are summarized in Table~\ref{DDHClass}. 
In terms of clinical decision support, great importance is given to the cutoff of $60^{\circ }$, distinguishing normal (type I) from dysplastic anatomy (type II) [13]. Results in Table \ref{DDHClass} indicate that our method can reduce the error rate from 10.121\% to 5.668\%. We observe that the misclassifications are mainly concentrated in the range of $58^{\circ}<\alpha<62^{\circ}$. In these cases, an angle estimate error can lead to misclassification, and how to deal with it remains an open issue. Fig. \ref{AngleComparison} shows the angle measurements by our method and the labels. Fig. \ref{AngleComparison} (a)-(d) show the precise automatic measurement results. There is only a minor difference between our method and the manual plot for landmarks and auxiliary measurement lines. Fig. \ref{AngleComparison} (e) and (f) are the cases with errors of more than 5 degrees. The image with alpha angle error larger than 5° is shown in Fig. \ref{AngleComparison} (e). It is caused by the offset of the bony rim to the inner side of the ilium. In Fig. \ref{AngleComparison} (f), one case with the beta angle error larger than 5 is shown. It is mainly caused by the positioning deviation of the labrum point. Although the angle errors are more than 5 degrees in these two cases, it is noted that they do not influence the visual judgment by experts, and thus their Graf classification will remain the same.

\subsection{Key Anatomical Structures Segmentation}
In this study, we make qualitative and quantitative comparisons on the segmentation of four key anatomical structures. Table \ref{DSCResult} and Table \ref{HDResult} quantitatively evaluate the effects of different network structures on the segmentation in terms of overall segmentation and edge accuracy, respectively.


As shown in Table \ref{DSCResult} and Table \ref{HDResult}, our method achieves comparable performance when compared with the SOTA segmentation networks. 
It can be observed in Table \ref{DSCResult} that DeeplabV3+ and our method outperform the Unet and PSPnet, with each of them has its strengths. In specific, DeeplabV3+ performs best in the segmentation of flat ilium and CO-Junction, while our method segment the lower limb and labrum with the best performance. Besides, compared with the original Mask R-CNN, adding SS Loss can improve the segmentation results, especially for small objects such as the lower limb. Moreover, with the SS Loss and the landmark branch being incorporated together, the segmentation results are further improved in both overall segmentation and edge accuracy evaluation. It indicates that jointly learning anatomical structures and landmarks can refine the predictions of the Mask R-CNN. Specifically, compared with the plain Mask R-CNN, our method increases the DSC of the four structures by 3.184\% on average and decreases the HD by an average of 12.706\%. These results demonstrate that the proposed multi-task learning network significantly improves the overall performance of segmentation and accuracy of borders.

\begin{table}
 \def\temptablewidth{1\columnwidth}
 \caption{Overall Segmentation Evaluation of Four Anatomical Structures by Dice Similarity Coefficient (DSC). The Asterisk (*) Denotes Significant Differences.}
\resizebox{\columnwidth}{!}{
    \begin{tabular*}{\temptablewidth}{@{\extracolsep{\fill}}lllll}
    \toprule[1pt]
    & Flat Ilium  & Lower Limb & Labrum & CO Junction \\
    &(std) & (std) &(std) & (std) \\
    \midrule            
    
    \multirow{2}[0]{*}{Unet} &  \multicolumn{1}{l}{0.882*} & \multicolumn{1}{l}{0.809*} & \multicolumn{1}{l}{0.791*} & \multicolumn{1}{l}{0.838*} \\
          & \multicolumn{1}{l}{(0.039)} & \multicolumn{1}{l}{(0.129)} & \multicolumn{1}{l}{(0.090)} & \multicolumn{1}{l}{(0.136)} \\
    \specialrule{0em}{1pt}{1pt}
    
    \multirow{2}[0]{*}{PSPnet} &  \multicolumn{1}{l}{0.867*} & \multicolumn{1}{l}{0.812*} & \multicolumn{1}{l}{0.820*} & \multicolumn{1}{l}{0.855*} \\
          & \multicolumn{1}{l}{(0.067)} & \multicolumn{1}{l}{(0.102)} & \multicolumn{1}{l}{(0.105)} & \multicolumn{1}{l}{(0.062)} \\
    \specialrule{0em}{1pt}{1pt}
    
    \multirow{2}[0]{*}{DeeplabV3+} &  \multicolumn{1}{l}{\textbf{0.893}} & \multicolumn{1}{l}{0.832*} & \multicolumn{1}{l}{0.840*} & \multicolumn{1}{l}{\textbf{0.873}} \\
          & \multicolumn{1}{l}{(0.080)} & \multicolumn{1}{l}{(0.117)} & \multicolumn{1}{l}{(0.099)} & \multicolumn{1}{l}{(0.084)} \\
    \specialrule{0em}{1pt}{1pt}
    
    \multirow{2}[0]{*}{Mask R-CNN} &  \multicolumn{1}{l}{0.869*} & \multicolumn{1}{l}{0.818*} & \multicolumn{1}{l}{0.817*} & \multicolumn{1}{l}{0.829*} \\
          & \multicolumn{1}{l}{(0.045)} & \multicolumn{1}{l}{(0.090)} & \multicolumn{1}{l}{(0.086)} & \multicolumn{1}{l}{(0.055)} \\
          
    \specialrule{0em}{1pt}{1pt}
    
    \multirow{2}[0]{*}{+ SS Loss} & \multicolumn{1}{l}{0.871*} & \multicolumn{1}{l}{0.827*} & \multicolumn{1}{l}{0.826*} & \multicolumn{1}{l}{0.846*} \\
          & \multicolumn{1}{l}{(0.044)} & \multicolumn{1}{l}{(0.077)} & \multicolumn{1}{l}{(0.078)} & \multicolumn{1}{l}{(0.055)} \\
    \specialrule{0em}{1pt}{1pt}    
    
    \multirow{2}[0]{*}{+ Landmark} & \multicolumn{1}{l}{0.876*} & \multicolumn{1}{l}{0.833*} & \multicolumn{1}{l}{0.829*} & \multicolumn{1}{l}{0.838*} \\
          & \multicolumn{1}{l}{(0.043)} & \multicolumn{1}{l}{(0.087)} & \multicolumn{1}{l}{(0.089)} & \multicolumn{1}{l}{(0.023)} \\
      \specialrule{0em}{1pt}{1pt}  
        
    \multirow{2}[1]{*}{Our Method} & \multicolumn{1}{l}{0.892} & \multicolumn{1}{l}{\textbf{0.838}} & \multicolumn{1}{l}{\textbf{0.841}} & \multicolumn{1}{l}{0.868} \\
         & \multicolumn{1}{l}{(0.039)} & \multicolumn{1}{l}{(0.087)} & \multicolumn{1}{l}{(0.065)} & \multicolumn{1}{l}{(0.043)} \\
    \bottomrule[1pt]
    \end{tabular*}%
    }
 \label{DSCResult}%
\end{table}%

\begin{table}
 \def\temptablewidth{1\columnwidth}
 \caption{Edge Accuracy Evaluation of Four Anatomical Structures by Hausdorff Distance (HD) (In Pixel). The Asterisk (*) Denotes Significant Differences.}
\resizebox{\columnwidth}{!}{
    \begin{tabular*}{\temptablewidth}{@{\extracolsep{\fill}}lllll}
    \toprule[1pt]
    & Flat Ilium  & Lower Limb & Labrum & CO Junction \\
    &(std) & (std) &(std) & (std) \\
    \midrule            
    \multirow{2}[0]{*}{Unet} &  \multicolumn{1}{l}{11.911*} & \multicolumn{1}{l}{16.204*} & \multicolumn{1}{l}{23.836*} & \multicolumn{1}{l}{22.214*} \\
          & \multicolumn{1}{l}{(21.060)} & \multicolumn{1}{l}{(10.984)} & \multicolumn{1}{l}{(16.549)} & \multicolumn{1}{l}{(25.319)} \\
          
    \specialrule{0em}{1pt}{1pt}
    
    \multirow{2}[0]{*}{PSPnet} &  \multicolumn{1}{l}{6.697*} & \multicolumn{1}{l}{8.199*} & \multicolumn{1}{l}{9.349*} & \multicolumn{1}{l}{20.516*} \\
          & \multicolumn{1}{l}{(10.279)} & \multicolumn{1}{l}{(13.537)} & \multicolumn{1}{l}{(10.769)} & \multicolumn{1}{l}{(25.742)} \\
          
    \specialrule{0em}{1pt}{1pt}
    
    \multirow{2}[0]{*}{DeeplabV3+} &  \multicolumn{1}{l}{\textbf{5.154}} & \multicolumn{1}{l}{5.861*} & \multicolumn{1}{l}{7.324*} & \multicolumn{1}{l}{15.857*} \\
          & \multicolumn{1}{l}{(4.248)} & \multicolumn{1}{l}{(4.395)} & \multicolumn{1}{l}{(4.150)} & \multicolumn{1}{l}{(18.501)} \\
          
    \specialrule{0em}{1pt}{1pt}
    
    \multirow{2}[0]{*}{Mask R-CNN} &  \multicolumn{1}{l}{21.273*} & \multicolumn{1}{l}{6.699*} & \multicolumn{1}{l}{7.566*} & \multicolumn{1}{l}{14.667*} \\
          & \multicolumn{1}{l}{(11.039)} & \multicolumn{1}{l}{(4.797)} & \multicolumn{1}{l}{(3.374)} & \multicolumn{1}{l}{(9.822)} \\
          
    \specialrule{0em}{1pt}{1pt}
    
    \multirow{2}[0]{*}{+ SS Loss} & \multicolumn{1}{l}{19.988*} & \multicolumn{1}{l}{6.289*} & \multicolumn{1}{l}{7.207*} & \multicolumn{1}{l}{13.786*} \\
          & \multicolumn{1}{l}{(11.085)} & \multicolumn{1}{l}{(3.302)} & \multicolumn{1}{l}{2.959)} & \multicolumn{1}{l}{(9.268)} \\
    \specialrule{0em}{1pt}{1pt}    
    
    \multirow{2}[0]{*}{+ Landmark} & \multicolumn{1}{l}{19.764*} & \multicolumn{1}{l}{6.080*} & \multicolumn{1}{l}{7.367*} & \multicolumn{1}{l}{14.180*} \\
          & \multicolumn{1}{l}{(11.163)} & \multicolumn{1}{l}{(4.331)} & \multicolumn{1}{l}{(3.001)} & \multicolumn{1}{l}{(10.747)} \\
      \specialrule{0em}{1pt}{1pt}  
        
    \multirow{2}[1]{*}{Our Method} & \multicolumn{1}{l}{19.214} & \multicolumn{1}{l}{\textbf{5.651}} & \multicolumn{1}{l}{\textbf{6.835}} & \multicolumn{1}{l}{\textbf{12.344}} \\
         & \multicolumn{1}{l}{(10.032)} & \multicolumn{1}{l}{(4.327)} & \multicolumn{1}{l}{(3.115)} & \multicolumn{1}{l}{(9.292)} \\
    \bottomrule[1pt]
    \end{tabular*}%
    }
 \label{HDResult}%
\end{table}

\begin{table}
  \centering
  \def\temptablewidth{1\columnwidth}
  \caption{Mean(and Standard Deviation) Distance of Three Landmarks. The Asterisk (*) Denotes Significant Differences.}
    \begin{tabular*}{\temptablewidth}{@{\extracolsep{\fill}}llll}
    \toprule[1pt]
     Mean Distance& Bony Rim & Lower Limb   & Midpoint of the \\
     (Pixel) & (STD) &  Point(STD) &  Labrum(STD) \\
    \midrule
      \multirow{2}[0]{*}{Unet} &  \multicolumn{1}{l}{7.765*} & \multicolumn{1}{l}{12.134*} & \multicolumn{1}{l}{7.239*} \\
                             & \multicolumn{1}{l}{(18.806)} & \multicolumn{1}{l}{(11.384)} & \multicolumn{1}{l}{(9.348)} \\
          
    \specialrule{0em}{2pt}{1pt}
    
     \multirow{2}[0]{*}{PSPnet} &  \multicolumn{1}{l}{10.187*} & \multicolumn{1}{l}{5.990*} & \multicolumn{1}{l}{5.308*} \\
                             & \multicolumn{1}{l}{(24.248)} & \multicolumn{1}{l}{(8.443)} & \multicolumn{1}{l}{(7.219)} \\
          
    \specialrule{0em}{2pt}{1pt}
    
     \multirow{2}[0]{*}{DeeplabV3+} &  \multicolumn{1}{l}{4.713*} & \multicolumn{1}{l}{5.092*} & \multicolumn{1}{l}{\textbf{4.480}} \\
                             & \multicolumn{1}{l}{(4.325)} & \multicolumn{1}{l}{(5.215)} & \multicolumn{1}{l}{(3.092)} \\
          
    \specialrule{0em}{2pt}{1pt}
     \multirow{2}[0]{*}{Mask R-CNN} &  \multicolumn{1}{l}{5.381*} & \multicolumn{1}{l}{5.887*} & \multicolumn{1}{l}{5.009*} \\
                             & \multicolumn{1}{l}{(3.297)} & \multicolumn{1}{l}{(5.882)} & \multicolumn{1}{l}{(3.478)} \\
          
    \specialrule{0em}{2pt}{1pt}
    
    \multirow{2}[0]{*}{+ SS Loss} & \multicolumn{1}{l}{5.256*} & \multicolumn{1}{l}{5.677*} & \multicolumn{1}{l}{4.635*}  \\
                                  & \multicolumn{1}{l}{(3.168)} & \multicolumn{1}{l}{(5.903)} & \multicolumn{1}{l}{(3.418)} \\
    \specialrule{0em}{2pt}{1pt}   
    
      \multirow{2}[0]{*}{-MASK +Landmark} & \multicolumn{1}{l}{6.087*} & \multicolumn{1}{l}{5.722*} & \multicolumn{1}{l}{5.108*}\\
                                & \multicolumn{1}{l}{(3.710)} & \multicolumn{1}{l}{(7.029)} & \multicolumn{1}{l}{(3.842)}  \\
      \specialrule{0em}{2pt}{1pt}  
    
    \multirow{2}[0]{*}{+ Landmark} & \multicolumn{1}{l}{4.926*} & \multicolumn{1}{l}{5.534*} & \multicolumn{1}{l}{4.897*}\\
                                & \multicolumn{1}{l}{(3.307)} & \multicolumn{1}{l}{(6.458)} & \multicolumn{1}{l}{(3.386)}  \\
      \specialrule{0em}{2pt}{1pt}  
        
    \multirow{2}[1]{*}{Our Method} & \multicolumn{1}{l}{\textbf{4.534}} & \multicolumn{1}{l}{\textbf{5.013}}& \multicolumn{1}{l}{4.563} \\
                              & \multicolumn{1}{l}{(3.024)} & \multicolumn{1}{l}{(5.803)} & \multicolumn{1}{l}{(3.201)}\\
    \bottomrule[1pt]
    \end{tabular*}%
 \label{PointResult}%
\end{table}%

\begin{figure}
\centerline{
    \begin{minipage}[b]{\columnwidth} 
    \centering
    \subfloat[]{
        \includegraphics[width=0.31\columnwidth]{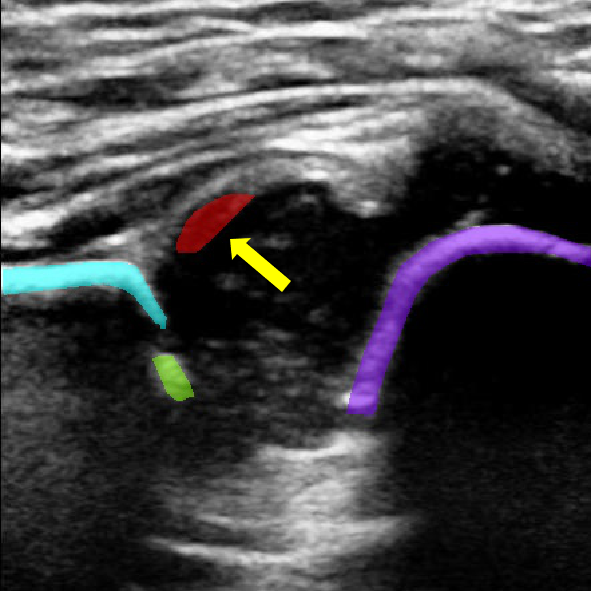} 
        \includegraphics[width=0.31\columnwidth]{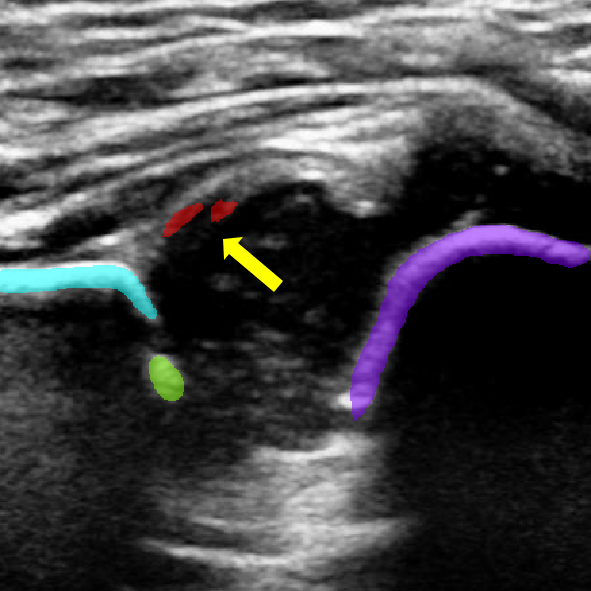} 
        \includegraphics[width=0.31\columnwidth]{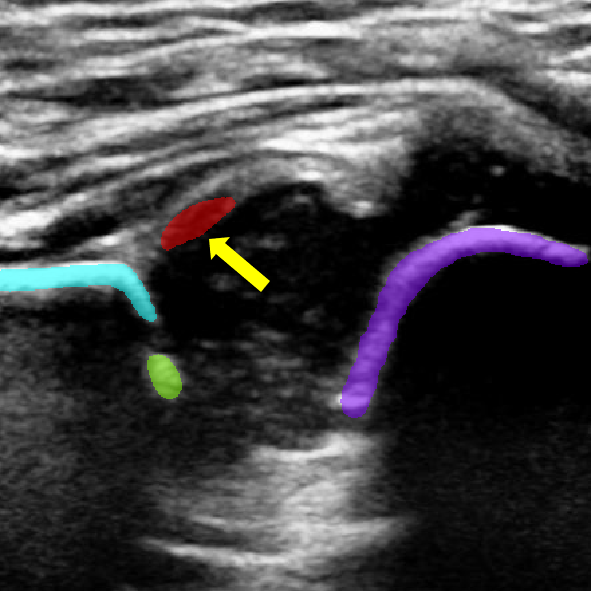}
        }\\
    \hspace{1mm}
    \subfloat[]{
        \includegraphics[width=0.31\columnwidth]{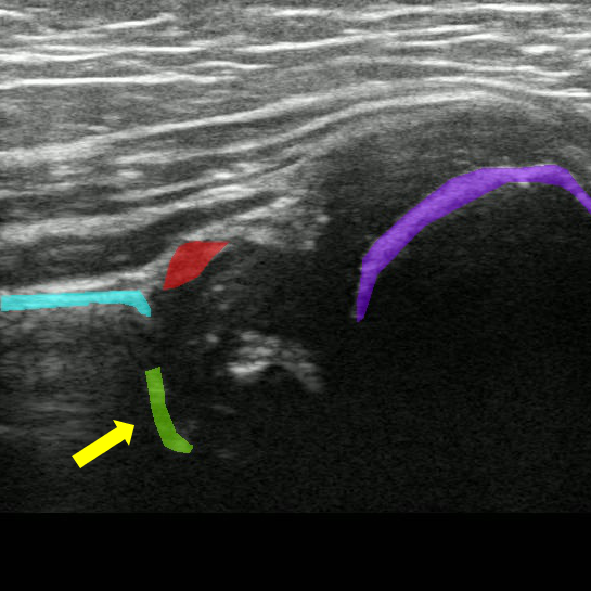} 
        \includegraphics[width=0.31\columnwidth]{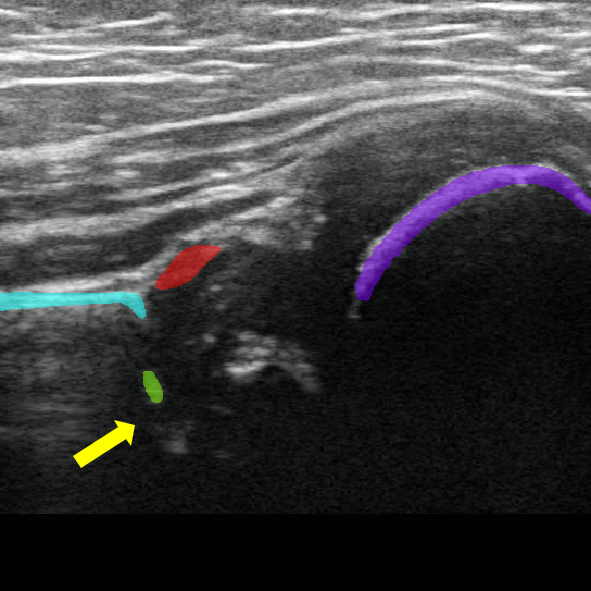} 
        \includegraphics[width=0.31\columnwidth]{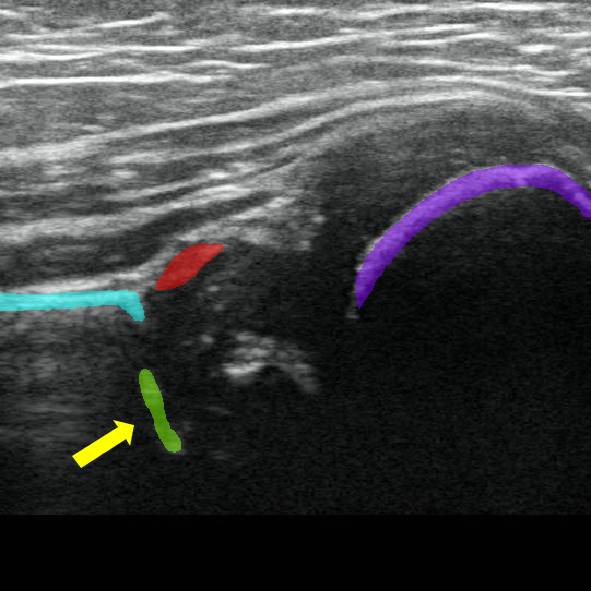}
        }\\
     \hspace{1mm}
    \subfloat[]{
        \includegraphics[width=0.31\columnwidth]{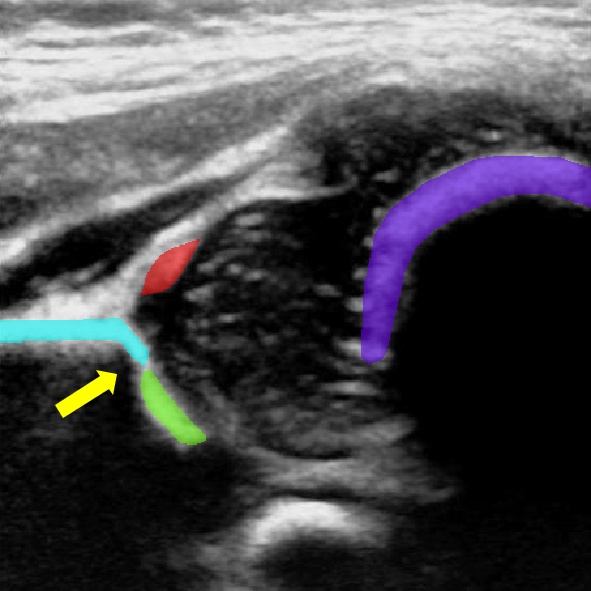} 
        \includegraphics[width=0.31\columnwidth]{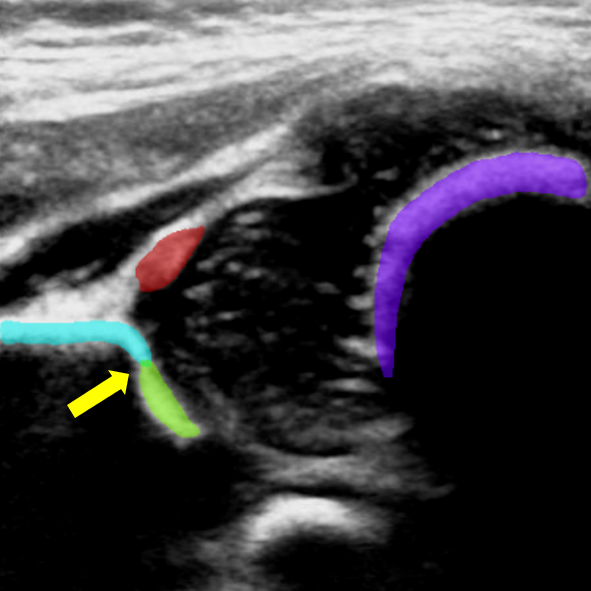} 
        \includegraphics[width=0.31\columnwidth]{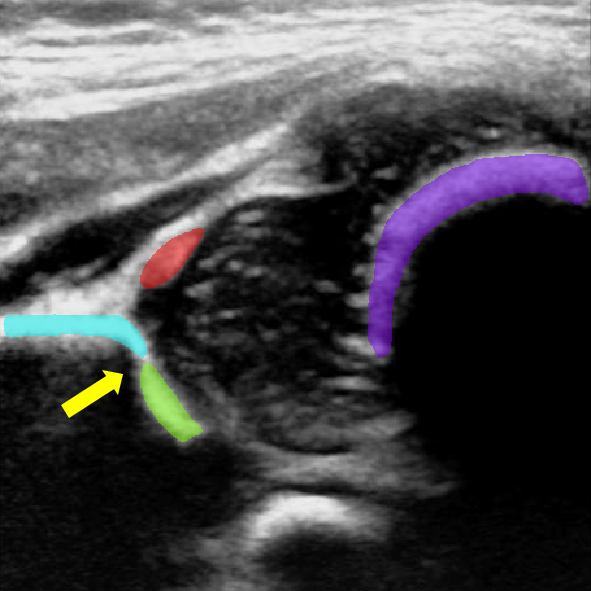}
        }\\
    \footnotesize \quad Ground Truth \quad\quad\quad\quad Mask R-CNN \quad\quad\quad\quad Our method
    \hspace{1mm}
    \end{minipage}}
   
\caption{Qualitative comparison among our method, Mask R-CNN, and ground truth(GT). (a) and (b) show two typical examples of the labrum and lower limb, respectively. The arrows indicate where the segmentation is improved. (c) For the challenging case in Fig. \ref{Challenges} (c), our method can correctly segment the flat ilium and the lower limb.}
\label{SegComparison}
\end{figure}

Fig. \ref{SegComparison} shows three examples of visual comparisons on the segmentation of the labrum and lower limb, respectively. From Fig. \ref{SegComparison} (a) and Fig. \ref{SegComparison} (b), we can see that the plain Mask R-CNN predicts incomplete segmentation results when the target is blurred due to the poor image quality and imperfect standard plane. In contrast, our method can well circumvent this problem by introducing structure priors, so that the model can learn the relationships among anatomical structures successfully. These findings verify the efficacy of our method in segmenting key anatomical structures. In clinical diagnosis, sometimes all bony parts of the acetabular roof are hyperechoic (see Fig. \ref{Challenges} (c)). The flat ilium is connected with the lower ilium, which may cause over-segmentation or under-segmentation, and subsequently, the bony rim is difficult to be distinguished. Thanks to the landmark detection branch and the bony rim loss, our method can learn the structural information of different parts and the two structures can be well separated, as shown in Fig. \ref{SegComparison} (c). In summary, the above results demonstrate that the proposed SS loss and BR loss are helpful to learn relationships among anatomical structures, predicting better segmentation results and bony rim locations.

\subsection{Landmark Detection}
Table \ref{PointResult} shows the mean and standard deviation distance of the predicted positions of three landmarks: the bony rim, the lower limb point, and the midpoint of the labrum. For the network without the landmark branch, the landmarks are determined only by the predicted mask, according to the location relationship among landmarks and anatomical structures as shown in Fig. \ref{DDHIntro}(a). 
As shown in Table \ref{PointResult}, our method achieved better performance than Unet, PSPnet, and DeeplabV3+ in localizing Bony Rim and Lower Limb Point.
It is noted that although DeeplabV3+ performs best in detecting the landmark of the midpoint of the labrum (4.480), our method obtains a very closed result (4.563). Furthermore, compared with the original Mask R-CNN, adding SS Loss and the landmark branch can reduce the detection errors. It can be observed in "-MASK+Landmark" and "Our Method" rows that our method performs better than the results with landmark branch only. Our method leads the baseline by an average of 0.722 pixels in terms of mean distance, which indicates the superiority of our proposed multi-task learning system.

\subsection{Influence of BR Loss}
For the sake of consistency in predicting the Bony Rim position, we propose the BR loss. The BR loss is added to the network with the landmark branch but no SS loss. The quantitative improvement is recorded in Table \ref{BRLossResul}. We evaluate it from the perspective of the segmentation of flat ilium, the localization error of bony rim, and the measurement error of beta angle since the BR loss only works on the bony rim and its related tasks. Experimental results show that BR loss improves the performance of multiple tasks. Especially in the measurement of beta angle, the error is reduced by 0.152°, and the relative error is reduced by 4.975\%. BR loss has a great contribution to the optimization of network performance. 

\begin{table}[!t]
  \centering
  \def\temptablewidth{1\columnwidth}
  \caption{Assessment of the Effects of BR Loss.}
    \begin{tabular*}{\temptablewidth}{@{\extracolsep{\fill}}lcccc}
    
    \toprule[1pt]
     & \multicolumn{2}{c}{Flat ilium} & Bony Rim error & \multicolumn{1}{c}{\multirow{2}[3]{*}{$\beta$ error (°)}} \\
\cmidrule{2-3}          & DSC   & HD (pixel) & (pixel) &  \\
    \midrule
     +BR Loss & $ \uparrow0.011 $ & $\downarrow$ 0.427  & $\downarrow$ 0.230 & $\downarrow$ 0.152 \\

    \bottomrule[1pt]
    \end{tabular*}%
     \label{BRLossResul}%
\end{table}%

\begin{table}[!t]
 \def\temptablewidth{1\columnwidth}
 \caption{The Effect of Different Regularization Weight Combinations on Model Performance.}
\resizebox{\columnwidth}{!}{
    \begin{tabular*}{\temptablewidth}{@{\extracolsep{\fill}}llllcc}
    \toprule[1pt]
    $\lambda$1  & $\lambda$2 & $\lambda$3 & $\lambda$4 & 
    \begin{tabular}[c]{@{}c@{}}Error of\\Angle $\alpha$ (°)\end{tabular}& 
    \begin{tabular}[c]{@{}c@{}}Error of\\Angle $\beta$ (°)\end{tabular}\\
    \midrule            
    \multirow{8}{*}{1} & \multirow{5}{*}{1} & \multirow{3}{*}{1} & 1 & 2.418 & \textbf{2.641} \\
                       &                    &                   & 0.5 & 2.358 & 2.942\\
                       &                    &                   & 0.1 & 2.028 & 3.019\\
                      \specialrule{0em}{1pt}{1pt}
                        \cline{3-6}
                        \specialrule{0em}{1pt}{1pt}
                       &                    & 0.5 & \multirow{2}{*}{1}& 2.291 & 2.748\\   
                       &                    & 0.1 &                   & 2.308 & 2.882\\  
                       \specialrule{0em}{1pt}{1pt}
                        \cline{2-6}
                        \specialrule{0em}{1pt}{1pt}
                        
                       & \multirow{2}{*}{0.5}& 0.5 & \multirow{2}{*}{1}& \textbf{2.011} & 2.899\\  
                       &                    & 0.1   &                  & 2.175 & 2.822\\  
                        \specialrule{0em}{1pt}{1pt}
                        \cline{2-6}
                        \specialrule{0em}{1pt}{1pt}
                       & 0.1              & 0.1   &        0.1       & 2.201 & 2.791\\  
    \specialrule{0em}{1pt}{1pt}
    \cline{1-6}
    \specialrule{0em}{1pt}{1pt}
         0.5           &  1                 & 1  &         0.5       & 2.582 & 3.016\\

    \bottomrule[1pt]
    \end{tabular*}%
    }
 \label{Weight}%
\end{table}

\begin{table}[!t]
  \centering
  \def\temptablewidth{1\columnwidth}
  \caption{Test Performance Comparison of Our Method and Mask R-CNN in Three Image Quality. }
 \resizebox{\columnwidth}{!}{
 \begin{tabular}{ccccccc}
     \toprule[1pt]
     method  & \begin{tabular}[c]{@{}c@{}}Image\\quality  \end{tabular} & DSC & \begin{tabular}[c]{@{}c@{}}HD\\(pixel)  \end{tabular}& 
    \begin{tabular}[c]{@{}c@{}}landmark \\(pixel)  \end{tabular}&\begin{tabular}[c]{@{}c@{}}Error of\\Angle $\alpha$ (°) \end{tabular} &\begin{tabular}[c]{@{}c@{}}Error of\\Angle $\beta$ (°)  \end{tabular}\\
     \midrule
     \multirow{3}{*}{Mask R-CNN}&good&  \multicolumn{1}{c}{0.838} & \multicolumn{1}{c}{12.568} & \multicolumn{1}{c}{5.221} & \multicolumn{1}{c}{2.527} & \multicolumn{1}{c}{2.758}\\
     &medium &\multicolumn{1}{c}{0.829} & \multicolumn{1}{c}{13.044} & \multicolumn{1}{c}{5.471} & \multicolumn{1}{c}{2.771} & \multicolumn{1}{c}{3.240} \\
     &poor & \multicolumn{1}{c}{0.832} & \multicolumn{1}{c}{12.042} & \multicolumn{1}{c}{5.585} & \multicolumn{1}{c}{2.982} & \multicolumn{1}{c}{3.662} \\
     \midrule
     \multirow{3}{*}{Our method}&good&  \multicolumn{1}{c}{0.884} & \multicolumn{1}{c}{10.869} & \multicolumn{1}{c}{3.852} & \multicolumn{1}{c}{1.807} & \multicolumn{1}{c}{2.413}\\
     &medium &\multicolumn{1}{c}{0.849} & \multicolumn{1}{c}{10.946} & \multicolumn{1}{c}{4.884} & \multicolumn{1}{c}{2.659} & \multicolumn{1}{c}{3.011} \\
     &poor & \multicolumn{1}{c}{0.847} & \multicolumn{1}{c}{11.218} & \multicolumn{1}{c}{5.374} & \multicolumn{1}{c}{2.197} & \multicolumn{1}{c}{3.274} \\
     \bottomrule[1pt]
\end{tabular}
}
\label{three_different_test}%
\end{table}%

\begin{table*}
  \centering
  \caption{Comparison Results of Two Kinds of Machines and Two Types of the Hip Joint.}
    \resizebox{\textwidth}{!}{
\begin{tabular}{cccccccccccccc}
        \toprule[1pt]
          & \multicolumn{4}{c}{DSC}   & \multicolumn{4}{c}{HD (pixel)}     & \multicolumn{3}{c}{Landmark (pixel)}   & \multicolumn{2}{c}{Angle error (°)} \\ 
          \specialrule{0em}{1pt}{1pt}   
          \cline{2-14} 
          \specialrule{0em}{1pt}{1pt}   
          & \begin{tabular}[c]{@{}c@{}}Flat\\ Ilium\end{tabular} & \begin{tabular}[c]{@{}c@{}}Lower\\ Limb\end{tabular} & Labrum & \begin{tabular}[c]{@{}c@{}}CO\\ Junction\end{tabular} & \begin{tabular}[c]{@{}c@{}}Flat\\ Ilium\end{tabular} & \begin{tabular}[c]{@{}c@{}}Lower\\ Limb\end{tabular} & Labrum & \begin{tabular}[c]{@{}c@{}}CO\\ Junction\end{tabular} & \begin{tabular}[c]{@{}c@{}}Bony\\ Rim\end{tabular} & \begin{tabular}[c]{@{}c@{}}Lower \\ Limb\\ Point\end{tabular} & \begin{tabular}[c]{@{}c@{}}Midpoint \\ of the \\ Labrum\end{tabular} & Alpha  & Beta  \\ 
        \midrule
            Machine 1 & 0.873  & 0.820  & 0.833 & 0.836  & 22.673   & 5.895  & 6.609  & 12.580  & 4.485  & 5.158 & 4.302  & 2.318  & 2.881  \\
            Machine 2 & 0.892  & 0.850  & 0.848 & 0.866  & 16.909   & 5.715  & 6.784  & 13.183  & 4.861  & 5.264  & 4.593 & 2.217 & 2.756  \\ 
        \midrule
            Type I    & 0.889   & 0.839   & 0.840   & 0.858   & 19.762    & 5.705  & 6.709  & 13.074  & 4.521  & 5.260  & 4.382 & 2.265   & 2.826  \\
            Type II   & 0.890   & 0.841   & 0.844  & 0.853   & 19.568    & 5.511  & 6.843  & 13.159  & 4.454  & 5.197  & 4.218 & 2.290  & 2.880\\
          \bottomrule[1pt]
\end{tabular}
}
 \label{Significantdiff}%
\end{table*}%

\begin{figure*}
\centerline{
    \includegraphics[width=0.9\textwidth]{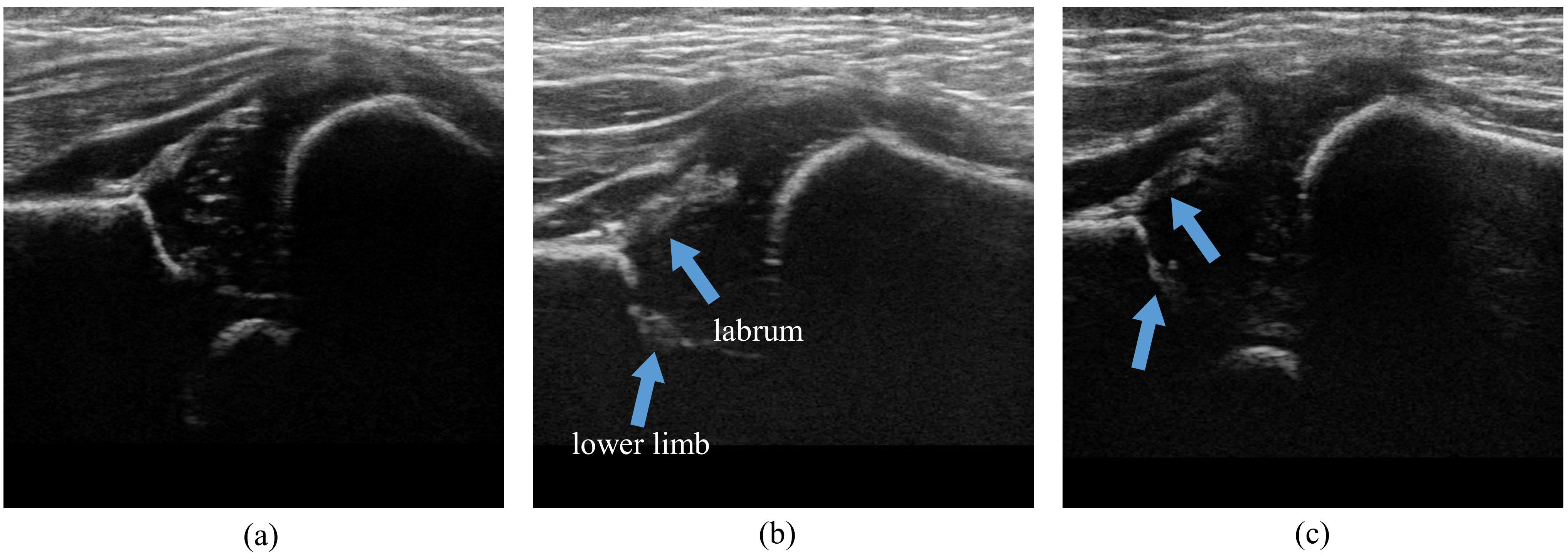}
    }
\caption{Three types of DDH images with different qualities. (a) good; (b) medium quality with blurred and unclear edges; (c) poor quality with invisible structures.}
\label{different_quality}
\end{figure*}

\begin{figure}
\centerline{
    \subfloat[]{\includegraphics[width=0.45\columnwidth]{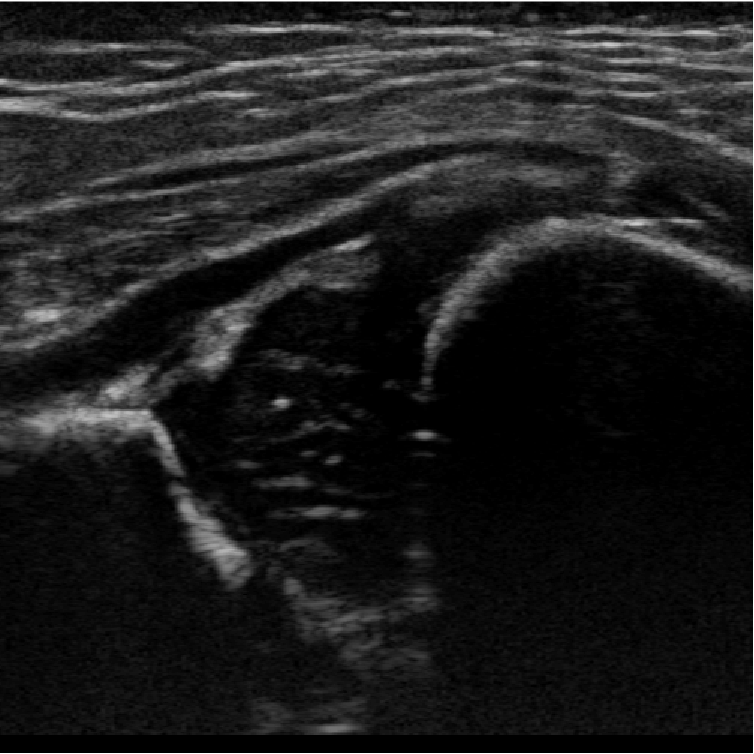}}
    \hspace{1mm}
    \subfloat[]{\includegraphics[width=0.45\columnwidth]{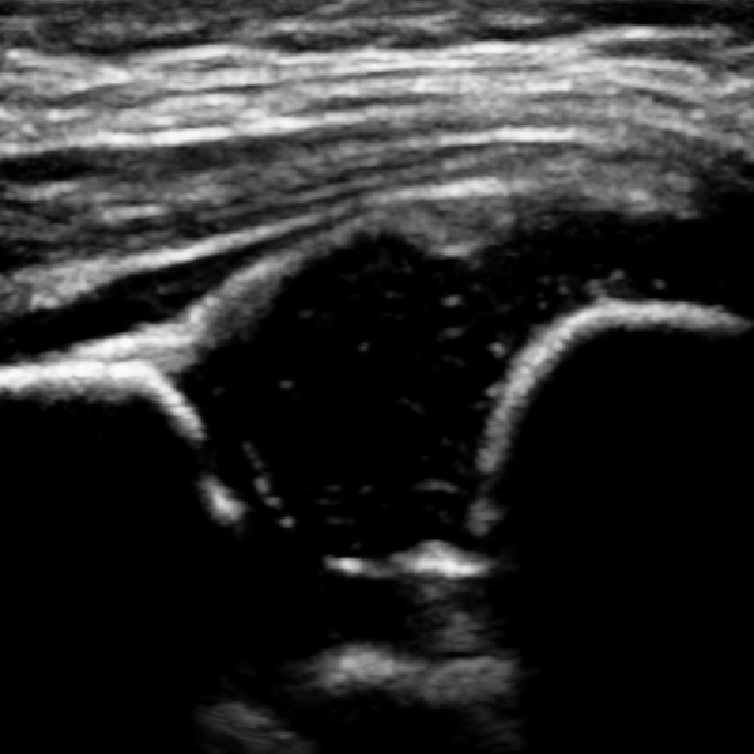}}
    }
\caption{Typical US images of two kinds of the machine in our dataset. (a) US image from Machine 1. (b) US image from Machine 2.}
\label{TwoMachines}
\end{figure}

\subsection{Influence of Regularization Weight}

There are four parameters to control the weight of losses in \ref{TotalLoss}. Our hypothesis is that, the shape similarity based SS loss is general and critical for the accurate segmentation of all the four anatomical structures in the DDH US image, It intuitively plays a more important role than the consistency based losses $L_{landmark}$ and $L_{BR}$. Therefore, we assigned 1.0 for $L_{SS}$, while assigned 0.5 for both $L_{landmark}$ and $L_{BR}$. 

We further conducted experiments to verify our hypothesis and presented the influences of the parameters. As shown in Table \ref{Weight}, we tried several typical combinations of the coefficients. When comparing the four parameters, larger $\lambda$4 often leads to better alpha angle results. While larger $\lambda$2 and $\lambda$3 often contribute to slightly better beta angle results. Because the alpha angle is the main indicator for DDH evaluation, we finally take the coefficient setting in our paper as a compromise for alpha and beta angles.

\subsection{Influence of Image Quality, Machines and Types}

We test the three types of images separately to check the stability and robustness of our method. According to the differences in image quality, the test data (247 images) are divided into three categories: good (123 images), medium (70 images), and poor (54 images), as exemplars shown in Fig.\ref{different_quality}. We tested our method on those three categories and showed the results in Table \ref{three_different_test}. As we can see, our proposed method presents a slight difference for images with different qualities. Good image quality contributes to better accuracy than poor images. When compared with the Mask R-CNN, our method shows superiority in keep high accuracy. This proves the advantages of our novel frameworks.

In addition, the dataset used in our experiment is from two types of US machines. To investigate whether our method performs significantly differently on two machines, we compare the experimental results in Table \ref{Significantdiff}. It can be seen in Fig. \ref{TwoMachines} that there is a visual difference between the two types of machine images. The image of Machine 2 is clearer than that of Machine 1. However, the p-values of the alpha angle and the beta angle by the two-sample t-test are 0.118 and 0.152, which shows that there is no significant difference between the two machines. The deviation caused by different image qualities can be eliminated in our method, especially in the measurement of alpha and beta angles. To a certain extent, it indicates that our method is not affected by image quality in angle measurement, and has good robustness.

Moreover, the dataset of hip joints is divided into two types, type I and type II. The average test data for each fold contains 215 type I cases and 32 type II cases. Taking into account the imbalance in the number of images of the two types, 
we train the model by using two types of data together but evaluate the test results separately.  As shown in Table \ref{Significantdiff}, the p-values of the alpha angle and the beta angle by the two-sample t-test are 0.802 and 0.523, which shows that there is no significant difference between errors of alpha and beta angles of the two types of images. Please note that, due to the lack of sufficient positive cases (type III and type IV), our method is only validated on type I and II images. In the future, we will collect more positive cases to verify the reliability of our proposed method.

\section{Discussion}

US screening is crucial in the early diagnosis of DDH, and the Graf method, which is commonly used clinically, requires a high demand for clinicians' technical level. Therefore, we propose a multi-task learning network for automatic DDH measurements to assist clinical diagnosis. Experiments prove that our method is accurate and robust, and has great value for clinical application.

There have been many studies on the computer-aided diagnosis of DDH limited to the segmentation of local anatomical structures in the past. Compared with our deep learning-based methods, the methods in \cite{kocer2016measuring,quader2017automatic,quader2015automatic} based on traditional machine learning are limited to complex feature extraction designs and smaller data sets. Other deep learning-based methods  \cite{zhang2014facial,zhang2018end,golan2016fully,hareendranathan2017toward,sezer2020deep} rely on the segmentation of key structures, such as ilium or lower limb, while our method combines the information of anatomical structures and multiple landmarks to improve the accuracy and robustness of angle measurement. Besides, unlike the method in \cite{quader2017automatic,quader2015automatic,golan2016fully}, which only focuses on the measurement of dysplasia metrics, we have added the identification of four key anatomical structures to ensure the reliability of the experiment and provide a reference for doctors to select the appropriate plane for diagnosis. In addition to the framework of multi-task learning, we also incorporate extra structure priors to improve accuracy and robustness. As for the challenge of the fragmental segmentation, we propose a shape similarity (SS) loss to regularize the shape of anatomical structures so that an intact mask can be estimated from an incomplete anatomical structure. We further propose a bony rim (BR) loss to enforce the bony rim estimated from the segmentation of flat ilium to be consistent with the detected landmark, to improve both structure segmentation and landmark detection.

In addition, our method has some limitations. First of all, the image style differences caused by different types of US machines may cause performance degradation of the network model in practical applications. Although there is no significant difference in automatic angle measurement between the two machine models in our experiment, it is not the case in segmentation. We will consider style transfer and other methods to deal with the issue in the future. The lack of positive cases is also a limitation of our method. On the image of positive cases, some structures, such as the labrum, may not be visible due to structural deformation of the hip. In this case, our method may have unexpected problems. Therefore, we will continue to collect more DDH cases in the future, so that our method can work well on any type of hip joint. We will also study more on other directions of DDH-assisted diagnoses, such as video sequence and three-dimensional volume data analysis.

\section{Conclusion}
In this paper, we propose a multi-task learning framework for automatic DDH measurement. Mask R-CNN with a landmark detection branch is adopted to effectively learn the relationships among anatomical structures and landmarks of the hip joint. By imposing two novel structure priors on landmark detection and structure segmentation, we show that the performance of the network can be significantly improved in terms of accuracy and robustness, particularly for US images with incomplete and touching anatomical structures. Experiments demonstrate that our method shows great potential for clinical application, with 93\% alpha and 85\% beta angle estimation errors less than 5 degrees.

\bibliographystyle{IEEEtran}
\bibliography{IEEEabrv,ref01}

\end{document}